\documentclass[letterpaper,10pt]{article}
\usepackage{amssymb,amsthm,amsmath}
\usepackage{amsfonts}
\usepackage{graphicx}
\usepackage{psfrag}
\usepackage{color}
\usepackage[all]{xy} 

\psfrag{Crate}{$r$}
\psfrag{Qrate}{$S$}
\psfrag{R}{$R$}
\psfrag{S}{$S$}

\DeclareMathOperator{\tr}{Tr}

\DeclareMathOperator{\E}{\mathbb{E}}

\begin{document}
\def\qedsymbol{\rule{7pt}{7pt}}
\def\supp{ {\rm{supp \,}}}
\def\dist{ {\rm{dist }}}
\def\dim{ {\rm{dim \,}}}
\def\oti{{\otimes}}
\def\bra#1{{\langle #1 |  }}
\def\lb{ \left[ }
\def\rb{ \right]  }
\def\tilde{\widetilde}
\def\bar{\overline}
\def\*{\star}
\def\({\left(}		\def\BL{\Bigr(}
\def\){\right)}		\def\BR{\Bigr)}
	\def\BBL{\lb}
	\def\BBR{\rb}
%

\def\1{{\mathbf{1} }}

\def\bb{{\bar{b} }}
\def\ab{{\bar{a} }}
\def\zb{{\bar{z} }}
\def\zbar{{\bar{z} }}
\def\inv#1{{1 \over #1}}
\def\half{{1 \over 2}}
\def\d{\partial}
\def\der#1{{\partial \over \partial #1}}
\def\dd#1#2{{\partial #1 \over \partial #2}}
\def\vev#1{\langle #1 \rangle}
\def\ket#1{ | #1 \rangle}
\def\rvac{\hbox{$\vert 0\rangle$}}
\def\lvac{\hbox{$\langle 0 \vert $}}
\def\2pi{\hbox{$2\pi i$}}
\def\e#1{{\rm e}^{^{\textstyle #1}}}
\def\grad#1{\,\nabla\!_{{#1}}\,}
\def\dsl{\raise.15ex\hbox{/}\kern-.57em\partial}
\def\Dsl{\,\raise.15ex\hbox{/}\mkern-.13.5mu D}
\def\b#1{\mathbf{#1}}
\newcommand{\proj}[1]{\ket{#1}\bra{#1}}
\def\braket#1#2{\langle #1 | #2 \rangle}
%
%
\def\th{\theta}		\def\Th{\Theta}
\def\ga{\gamma}		\def\Ga{\Gamma}
\def\be{\beta}
\def\al{\alpha}
\def\ep{\epsilon}
\def\vep{\varepsilon}
\def\la{\lambda}	\def\La{\Lambda}
\def\de{\delta}		\def\De{\Delta}
\def\om{\omega}		\def\Om{\Omega}
\def\sig{\sigma}	\def\Sig{\Sigma}
\def\vphi{\varphi}
%
%
\def\CA{{\cal A}}	\def\CB{{\cal B}}	\def\CC{{\cal C}}
\def\CD{{\cal D}}	\def\CE{{\cal E}}	\def\CF{{\cal F}}
\def\CG{{\cal G}}	\def\CH{{\cal H}}	\def\CI{{\cal J}}
\def\CJ{{\cal J}}	\def\CK{{\cal K}}	\def\CL{{\cal L}}

\def\CM{{\cal M}}	\def\CN{{\cal N}}	\def\CO{{\cal O}}
\def\CP{{\cal P}}	\def\CQ{{\cal Q}}	\def\CR{{\cal R}}
\def\CS{{\cal S}}	\def\CT{{\cal T}}	\def\CU{{\cal U}}
\def\CV{{\cal V}}	\def\CW{{\cal W}}	\def\CX{{\cal X}}
\def\CY{{\cal Y}}	\def\CZ{{\cal Z}}

\def\rvac{\hbox{$\vert 0\rangle$}}
\def\lvac{\hbox{$\langle 0 \vert $}}
\def\comm#1#2{ \BBL\ #1\ ,\ #2 \BBR }
\def\2pi{\hbox{$2\pi i$}}
\def\e#1{{\rm e}^{^{\textstyle #1}}}
\def\grad#1{\,\nabla\!_{{#1}}\,}
\def\dsl{\raise.15ex\hbox{/}\kern-.57em\partial}
\def\Dsl{\,\raise.15ex\hbox{/}\mkern-.13.5mu D}
\def\beq{\begin {equation}}
\def\eeq{\end {equation}}
\def\to{\rightarrow}
\def\h#1{\widehat{#1}}
\newtheorem{lem}{Lemma}
\newtheorem{prop}{Proposition}
\newtheorem{theo}{Theorem}
\newtheorem{dfn}{Definition}
\newtheorem{cor}{Corollary}

\def\diag{\mbox{diag}}

\def\argmax{\mbox{argmax}}
\definecolor{gray}{gray}{.9}
\def\com#1{\vspace{.1in}\fcolorbox{black}{gray}{\begin{minipage}{5.5in}#1\end{minipage}}\vspace{.1in}}
\title{Capacity Theorems for Quantum Multiple Access Channels: \\
Classical-Quantum and Quantum-Quantum \\ Capacity Regions}
\author{
Jon Yard\\
\it{Department of Electrical Engineering, Stanford University,} \\
\it{Stanford, California 94305, USA} \\
\\
Igor Devetak \\
\it{Electrical Engineering Department, University of Southern California,} \\ 
\it{Los Angeles, California 90089, USA}\\
\\
Patrick Hayden \\
\it{School of Computer Science, McGill University,}\\ 
\it{ Montreal, QC H3A 2A7 Canada}\\
}
\date{August 3, 2005}
\maketitle

\begin{abstract}
We consider quantum channels with two senders and one receiver.  For an arbitrary such channel, we give multi-letter characterizations of two different two-dimensional capacity regions.  The first region is comprised of the rates at which it is possible for one sender to send classical information, while the other sends quantum information.  The second region consists of the rates at which each sender can send quantum information.  For each region, we give an example of a channel for which the corresponding region has a single-letter description.  One of our examples relies on a new result proved here, perhaps of independent interest, stating that the coherent information over any degradable channel is concave in the input density operator.  We conclude with connections to other work and a discussion on generalizations where each user simultaneously sends classical and quantum information. 
\end{abstract}

\section{Introduction} \label{section:introduction}
A classical multiple access channel with two senders and one receiver is described by
a probability transition matrix $p(z|x,y)$.
For the situation in which each sender wishes to send independent information, Ahlswede \cite{ahlswede} and Liao \cite{liao} showed that the capacity region $\CC$ admits a single-letter characterization, given by the convex hull of the closure of the set of rate pairs $(r,s)$ satisfying
\begin{eqnarray*}
r &<& I(X;Z|Y) \\
s &<& I(Y;Z|X) \\
r+s &<&I(XY;Z)
\end{eqnarray*}
for some $p(x)p(y).$
Further analysis by Cover, El Gamal and Salehi
\cite{ces} gives single-letter characterizations of a set of correlated sources which can be reliably transmitted over a
multiple access channel,
generalizing the above, as well as
Slepian-Wolf source coding and cooperative multiple access channel capacity.  They also give a multi-letter expression for the capacity region, showing that an i.i.d.\ source $(U,V)$ can be reliably transmitted if and only if
\begin{eqnarray*}
H(U|V) &<& \frac{1}{n} I(X^n;Z^n|U^nY^n) \\
H(V|U) &<& \frac{1}{n} I(Y^n;Z^n|V^nX^n) \\
H(U,V) &<& \frac{1}{n} I(X^nY^n;Z^n)
\end{eqnarray*}
for some $n$ and $p(x^n|u^n),p(y^n|v^n)$, where $x^n$ refers to the sequence of symbols $(x_1,\dotsc,x_n)$.  A similar convention has been used for sequences of jointly distributed random variables, as
$X^n\equiv (X_1,\dotsc,X_n)$.  Such a characterization is of limited practical use, however, as it does not apparently lead to a finite computation for deciding if a source can be transmitted.

In quantum Shannon theory, various capacities of a single quantum channel are not currently known to be computable in general. It is known 
\cite{ss} that the capacity of an arbitrary quantum channel for transmitting quantum information cannot be expressed as a single-letter optimization problem.  Furthermore, the classical capacity of a quantum channel is only known to be additive in this sense when the encoder is restricted to preparing product states.

Winter \cite{wintermac} has shown that the capacity region of a multiple access channel with classical inputs and a quantum output for the transmission of independent classical messages admits a single letter characterization which is identical in form to that of $\CC$.  Results on the classical capacity region of quantum binary adder channels are contained in \cite{gleb,kw}.
In what follows, we will examine the capacity region of an arbitrary quantum multiple access channel with quantum inputs and a quantum output, used in two distinct ways for the transmission of uncorrelated information from each terminal.  Our first result describes the capacity region for the case in which one user sends quantum information, and the other classical.  The second result characterizes the capacity region for the scenario in which each user wishes to send only quantum information.

The paper is organized as follows.  Section~\ref{section:background} contains the relevant background material necessary to state and prove our main results.  This includes mention of the notational conventions we will use throughout the paper, definitions of the distance measures for states we will use, as well as definitions of the information quantities which will characterize our rate regions.  We also introduce two of the three equivalent information processing tasks that will be considered in this paper, entanglement transmission and entanglement generation.  Section~\ref{section:mainresults} contains statements of Theorems 1 and 2, the main results of this paper.  We collect various relationships between our distance measures, a number of lemmas, and statements of existing coding propositions in Section~\ref{section:proofs}, which also contains the proofs of Theorems 1 and 2.  In Section~\ref{section:sstequiv}, a third information processing scenario, strong subspace transmission, will be introduced.  All three scenarios will be proved equivalent in that section as well.  Section~\ref{section:discussion} relates results contained in this paper to existing and future results.  For each of the main theorems, the appendix gives an example of a quantum multiple access channel for which the associated description is single-letter, in the sense that it has a characterization in terms of an optimization of single-letter information quantities (Sections~\ref{section:appendix:cqadd} and \ref{section:appendix:qqadd}).
Proofs of a new concavity result for degradable channels (Section~\ref{section:appendix:concavity}), of the convexity of our capacity regions (Section~\ref{section:appendix:convex}), and of the sufficiency of the bound on the cardinality of the set of classical message states for cq protocols (Section~\ref{section:appendix:cardinality}) appear there as well.

\section{Background} \label{section:background}
A typical quantum system will be labeled $A$.  Its Hilbert space will be $\CH_A$. The dimension of $\CH_A$ will be abbreviated as $|A| = \dim\CH_A$. For convenience, the label $A$ will often be shorthand for some collection of operators on $\CH_A$ when the context makes this apparent.  For example, a density matrix $\rho\in A$ refers to a normalized, positive operator $\rho\colon \CH_A\rightarrow \CH_A$.  We will often abbreviate this by writing $\rho^A$ to remind the reader of the system to which $\rho$ belongs.  Saying that $\CN\colon A\rightarrow B$ is a channel will really mean that $\CN\colon \CB(\CH_A) \rightarrow \CB(\CH_B)$ is completely-positive and trace-preserving.   Two systems $A$ and $B$ may be combined with a tensor product, resulting in the system $AB$, where $\CH_{AB} \equiv \CH_A\otimes \CH_B.$ The system $A^n$ has a Hilbert space
$\CH_{A^n} \equiv \CH_A^{\otimes n},$ and the various operator algebras described by $A^n$ will be appropriate subsets of
$\CB(\CH_A^{\otimes n})$.  
We will freely identify $\CN \equiv \CN\otimes 1^C,$ where $C$ is any other system, in order to simplify long expressions.  This procedure will always result in a unique completely positive map, since every channel in this paper will be completely positive.
The maximally mixed state on a Hilbert space $\CH_A$ will always be written as $\pi^A = \frac{1}{|A|}1^A,$ and we reserve the symbol $\ket{\Phi}$ for bipartite states which are maximally entangled.  An exception to this convention will be made when, given a density matrix $\rho^{A'}$, we write $\ket{\Phi_\rho}^{AA'}$ for a purification of $\rho^{A'}.$  When we write the density matrix of a pure state $\ket{\psi},$ we will freely make the abbreviation
$\psi \equiv \proj{\psi}.$

We will use the following conventions for distance measures between states.  If $\rho$ and $\sigma$ are density matrices, we will write
\[F(\sigma,\rho) = \left(\tr\sqrt{\sqrt{\rho} \sigma \sqrt{\rho}}\right)^2\]
for (the squared version of) the fidelity \cite{uhl}.  It is not hard to check that $F$ is symmetric.  For two pure states, this reduces to
\[F(\ket{\phi},\ket{\psi}) = |\braket{\phi}{\psi}|^2,\]
while for a pure state and a mixed state,
\[F(\ket{\phi},\rho) = \bra{\phi}\rho\ket{\phi} = \tr\phi\rho.\]
In this last case, we may interpret the fidelity as the success probability for a measurement which tests for the presence of the pure state $\phi$, when a physical system with density matrix $\rho$ is presented.  Indeed, for a POVM $\{\phi, 1-\phi\}$,
\[\Pr\{\text{measure }\phi | \text{ prepared }\rho\} = \tr\phi\rho = F(\phi,\rho).\]

The \emph{trace norm} of an operator $A\in \CB(\CH)$ is defined as the sum of its singular values, and can be expressed as
\[|A|_1 = \tr\sqrt{A^\dagger A}.\]
This gives rise to another useful distance measure on states, the
\emph{trace distance}, defined as the trace norm of the difference between the states. It can be written explicitly as
\[|\sigma - \rho|_1 = \tr\sqrt{(\sigma - \rho)^2},\]
and carries a normalization which assigns a distance of 2 to states with orthogonal support.

In order to introduce the information quantities which will be used to characterize our capacity regions, we first introduce the concept of a classical-quantum (cq) density matrix or state.  Let $\CX$ be a finite set and let $X$ be an $\CX$-valued random variable, distributed according to $p(x)$.  We can define a Hilbert space $\CH_X$ with a fixed orthonormal basis $\{\ket{x}^X\}_{x\in\CX}$, labeled by elements of the set $\CX$.  This sets up an identification
$\ket{\cdot}^X\colon\CX \rightarrow\CH_X$ between the elements of $\CX$ and that particular basis.  By this correspondence, the probability distribution $p(x)$ can be mapped to a density matrix $\rho = \sum_{x\in\CX}p(x)\proj{x}$
which is diagonal in the basis $\{\ket{x}\}_{x\in\CX}.$  Further, to every subset $S\subseteq \CX$ corresponds a projection matrix $\Pi_S = \sum_{x\in S}\proj{x}$ which commutes with $\rho$.
This way, we can express concepts from classical probability theory in the language of quantum probability, such as the equivalence $\Pr\{X\in S\} = \tr \rho \Pi_S$.  From the early development of quantum mechanics, noncommutativity has been seen to be the hallmark of quantum behavior.  It is to be expected that classical probability, embedded in quantum theory's framework, is described entirely with commuting matrices.

Consider now a collection of density matrices $\big\{\sigma^A_x\big\}_{x\in\CX},$ indexed by the finite set $\CX$.  If those states occur according to the probability distribution $p(x)$, we may speak of an \emph{ensemble} $\big\{p(x),\sigma^A_x\big\}$ of quantum states.  In order to treat classical and quantum probabilities in the same framework, a joint density matrix can be constructed
\[\sigma^{XA} = \sum_{x\in\CX}p(x)\proj{x}^X\otimes\sigma^A_x.\]
This is known as a \emph{cq state}, and describes the classical and quantum aspects of the ensemble on the \emph{extended Hilbert space} $\CH_X\otimes\CH_A$ \cite{dcr}.  The semiclassical nature of the ensemble is reflected in the embedding of a direct sum of Hilbert spaces $\bigoplus_{x\in\CX}\CH_{A_x}$ into $\CH_X\otimes\CH_A$.  This should be compared with the purely classical case, where a direct sum of one-dimensional vector spaces $\bigoplus_{x\in\CX}\mathbb{C}$ was embedded into $\CH_X$.
Just as our classical density matrix $\rho$ was diagonal in a basis corresponding to elements of $\CX$, the cq density matrix $\sigma$ is \emph{block-diagonal}, where the diagonal block corresponding to $x$ contains the non-normalized density matrix $p(x)\sigma_x$.  The classical state is recoverable as
$\rho = \tr_A \sigma,$ while the average quantum state is $\tr_X\sigma = \sum_{x\in\CX}\sigma_x$.  We will further speak of cqq states, which consist of two quantum parts and one classical.  When even more systems are involved, we will defer to the terminology cq to mean that some subsystems are classical, while some are quantum.  Such states are not only of interest in their own right; information quantities evaluated on cq states play an important role in characterizing what is possible in quantum information theory.
Now, let $\sigma$ be some cqq state, in block-diagonal form
\[\sigma^{ABX} = \sum_x p(x) \proj{x}^X\otimes\sigma_x^{AB}.\]
We write
\[H(A)_\sigma = H(\sigma^A) = -\tr \sigma^A\log\sigma^A\]
for the von Neumann entropy of the density matrix associated with $A$,
where $\sigma^A = \tr_{BX}\sigma$.
$H(AB)_\sigma$ is defined analogously.  We will omit subscripts when the state under consideration is apparent.
The \emph{mutual information} is defined as
\[I(X;B) \equiv H(X) + H(B) - H(XB).\]
Depending on the context,
the \emph{coherent information} \cite{sn96} will expressed in one of two ways.
For a fixed joint state $\sigma$, we write
\[I_c(A\,\rangle B) \equiv H(B) - H(AB) = -H(A|B).\]
Otherwise, if we are given a density matrix $\rho^{A'}$ and a channel $\CN\colon A'\rightarrow B$ which give rise to a joint state 
$(1^A\otimes \CN)(\Phi_\rho)$,
where $\ket{\Phi_\rho}^{AA'}$ is any purification of $\rho$, we will often use the notation
\[I_c(A\,\rangle B) = I_c(\rho,\CN) = H(\CN(\rho)) - H((1\otimes\CN)(\Phi_\rho)).\]  
It can be shown that this latter expression is independent of the particular purification $\ket{\Phi_\rho}$ that is chosen for $\rho$.

Despite their distinct forms, the mutual and coherent informations do share a common feature.  For a fixed input state, each is a convex function of the channel.
We further remark that the quantity
$I_c(A\,\rangle BX)$ can be considered as a conditional, or expected,
coherent information, as
\[I_c(A\,\rangle BX)_\omega = \sum_x p(x) I(A\,\rangle B)_{\omega_x}.\]
A particular departure of this quantity from its classical analog, the conditional mutual information $I(X;Y|Z)$, is that the latter is only equal to $I(X;YZ)$ when $X$ and $Z$ are independent, while the former always allows either interpretation, provided the conditioning variable is classical.  

Conditional coherent information arises in another context; suppose that $\boldsymbol{\CN}:A'\rightarrow XB$ is a \emph{quantum instrument} \cite{davies}, meaning that $\boldsymbol{\CN}$ acts as 
\[\boldsymbol{\CN}\colon\tau \mapsto \sum_x \proj{x}^X \otimes \CN_x(\tau).\]
The completely positive maps $\{\CN_x\}$ are the \emph{components} of the instrument.  While they are generally trace reducing, their sum $\CN = \sum_x \CN_x$ is always trace preserving.  It is not difficult to show that 
\[I_c(\rho,\boldsymbol{\CN}) = I_c(A\,\rangle BX),\]
where the latter quantity is computed with respect to to the state
\[\sum_x p(x) \proj{x}^X \otimes (1^A \otimes \CN_x)(\Phi_\rho^{AA'}).\]

For us, a \emph{quantum multiple-access channel} is a channel
$\CN\colon A'B'\rightarrow C$ with two inputs and one output.  We will assume that the inputs $A'$ and $B'$ are under the control of Alice and Bob, respectively, and that the output $C$ is maintained by Charlie.  We will present three different quantum information processing scenarios which, as we will see, lead to equivalent cq and qq capacity regions.

\paragraph{Classical-Quantum (cq) protocols}
These protocols will be relevant to Theorem~\ref{theo:cq} below.  Using a large number $n$ of instances of $\CN,$ Alice tries to send classical information to Charlie at rate $r$, while Bob simultaneously attempts  to convey \emph{quantum information} at rate $S$.  Alice's communication is in the sense that she tries to send Charlie one of $2^{nr}$ equiprobable classical messages, represented by the uniformly distributed random variable $M$.  To this end, we allow her to prepare arbitrary pure states $\ket{\phi_m}^{A'^n}$ at her input $A'^n$ to the channel.  It is assumed that neither Alice nor Bob shares any additional resources with Charlie or among themselves, such as entanglement or noiseless quantum channels.  We consider three different information processing tasks which Bob can perform, introduced in order of apparently increasing strength.
The first two, entanglement generation and entanglement transmission, are outlined below, as each plays an essential role in the proof of our main result.  The third, strong subspace transmission, is described in Section~\ref{section:sstequiv:sst}.  While not essential for the understanding of our main results, we include it in this paper because the composability properties implied by its more stringent constraints on successful communication make it particularly attractive as a building block for creating more intricate protocols from simpler ones.   That each of these aforementioned scenarios can justifiably be considered as ``sending quantum information" to Charlie will be proved in Sections~\ref{section:sstequiv:eteg} and \ref{section:sstequiv:etsst}, where we will show that each gives rise to the same collection of achievable rates.  

\subparagraph{I - Entanglement generation}
With the goal of eventually sharing near maximal entanglement with Charlie, 
Bob begins by preparing a bipartite pure state
$\ket{\Upsilon}^{BB'^n},$ entangled between a physical system $B$ located in his laboratory, and the $B'^n$ part of the inputs of $\CN^{\otimes n}$.  
Charlie's post-processing procedure will be modeled by a quantum instrument.  While the outer bound provided by our converse theorem will apply to any decoding modeled by an instrument, our achievability proof will require a less general approach, consisting of the following steps.

In order to ascertain Alice's message $M$, Charlie first performs some measurement on $C^n$ whose statistics are given by a POVM $\{\Lambda_{m}\}_{m\in 2^{nr}}$.  We let the result of that measurement be denoted $\widehat{M},$ his declaration of the message sent by Alice.  Based on the result of that measurement, he will perform one of $2^{nr}$ decoding operations $\CD'_m\colon C^n\rightarrow \widehat{B}.$  These two steps can be mathematically combined to define a \emph{quantum instrument} $\boldsymbol{\CD}\colon C^n\rightarrow \widehat{M}\widehat{B}$ with (trace-reducing) components
\[\CD_m\colon \tau\mapsto\CD'_m(\sqrt{\Lambda_m}\tau\sqrt{\Lambda_m}).\]
The instrument acts as
\[\boldsymbol{\CD}\colon \tau\mapsto \sum_{m=1}^{2^{nr}}
\proj{m}^{\widehat{M}}\otimes\CD_m(\tau),\]
and induces the trace preserving map $\CD\colon C^n\rightarrow \widehat{B}$, acting according to
\[\CD\colon \tau\mapsto\tr_{\widehat{M}}\boldsymbol{\CD}(\tau) = \sum_{m=1}^{2^{nr}}\CD_m(\tau).\]
We again remark that this is the most general decoding procedure required of Charlie.  Any situation in which he were to iterate the above steps by measuring, manipulating, measuring again, and so on, is asymptotically just as good as a single instance of the above mentioned protocol.
$(\{\ket{\phi_m}\}_{m\in 2^{nr}},\ket{\Upsilon}^{BB'^n},\boldsymbol{\CD})$ will be called a
$(2^{nr},2^{nS},n,\epsilon)$ \emph{cq entanglement generation code} for the channel $\CN$ if 
\begin{eqnarray}
2^{-nr}\sum_{m=1}^{2^{nr}} P_s^{\text{I}}(m,\ket{\Upsilon}) \geq 1-\epsilon, \label{cqIs}
\end{eqnarray}
where
\begin{eqnarray}
P_s^{\text{I}}(m,\ket{\Upsilon}) = F\left(\ket{m}\ket{\Phi}^{B\widehat{B}},
\boldsymbol{\CD}\circ\CN^{\otimes n}
(\phi_m^{A'^n}\otimes\Upsilon^{BB'^n})\right). \label{cqIPs}
\end{eqnarray}

We will say that $(r,S)$ is an \emph{achievable cq rate pair for entanglement generation} if there exists a sequence of $(2^{nr},2^{nS},n,\epsilon_n)$ cq entanglement generation codes
with $\epsilon_n\rightarrow 0$.
The capacity region $\CC\CQ_{\text{I}}(\CN)$ is defined to be the closure of the collection of all achievable cq rate pairs for entanglement generation.

\subparagraph{II - Entanglement transmission}
In this scenario, rather than generating entanglement with Charlie, Bob will act to transmit \emph{preexisting} entanglement to him.  We assume that 
Bob is presented with the $\tilde{B}$ part of the maximally entangled state $\ket{\Phi}^{B\tilde B}.$  It is assumed that he has complete control over $\tilde{B}$, while he has no access to $B$.  He will perform a physical operation in order to transfer the quantum information embodied in his system $\tilde{B}$ to the inputs
$B'^n$ of the channel, modeled by an encoding operation $\CE\colon \tilde{B}\rightarrow B'^n$.  The goal of this encoding will be to make it possible for Charlie, via post-processing of the information embodied in the system $C^n$, to hold the $\widehat{B}$ part of a state which is close to that which would have resulted if Bob had sent his system through a perfect quantum channel $\text{id}\colon\tilde{B}\rightarrow \widehat{B}$.  Here, we imagine that $\tilde{B}$ and $\widehat{B}$ denote two distinct physical systems with the same number of quantum degrees of freedom.  The role of the identity channel is to set up a unitary correspondence, or isomorphism, between the degrees of freedom of $\tilde{B}$ in Bob's laboratory and those of $\widehat{B}$ in Charlie's.  We will often tacitly assume that such an identity map has been specified ahead of time in order to judge how successful an imperfect
quantum transmission has been.  This convention will be taken for granted many times throughout the paper, wherein specification of an arbitrary state $\ket{\Psi}^{B\tilde{B}}$ will immediately imply specification of the state $\ket{\Psi}^{B\widehat{B}} = (1^B\otimes\text{id})\ket{\Psi}^{B\tilde{B}}.$
Decoding is the same as it is for scenario I.  

$(\{\ket{\phi_m}\}_{m\in 2^{nr}},\CE,\boldsymbol{\CD})$ will be called a $(2^{nr},2^{nS},n,\epsilon)$ \emph{cq entanglement transmission code} for the channel $\CN$ if
\begin{eqnarray}
 2^{-nr}\sum_{m=1}^{2^{nr}} P_s^{\text{II}}(m) \geq 1-\epsilon, \label{cqIIs}
\end{eqnarray}
where
\begin{eqnarray}
P_s^{\text{II}}(m) = F\left(\ket{m}\ket{\Phi}^{B\widehat{B}},
\boldsymbol{\CD}\circ\CN^{\otimes n}
(\phi_m^{A'^n}\otimes\CE(\Phi^{B\tilde{B}})\right). \label{cqIIPs}
\end{eqnarray}
Achievable rate pairs and the capacity region $\CC\CQ_{\text{II}}(\CN)$ are defined analogous to those for scenario~I.

Scenario III will be introduced in Section~\ref{section:sstequiv}, where it will also be shown that all three scenarios gives rise to the same set of achievable rates.  For this reason, we will henceforth only speak of a single capacity region
\[\CC\CQ(\CN) = \CC\CQ_{\text{I}}(\CN) =  \CC\CQ_{\text{II}}(\CN) =  \CC\CQ_{\text{III}}(\CN).\]

\paragraph{Quantum-Quantum protocols}
The subject of Theorem~\ref{theo:qq}, these protocols concern the case in which Alice and Bob each wish to send only quantum information to Charlie at rates $R$ and $S$, respectively.
As in the cq case, we will initially describe two different senses in which such a task can be considered.  Again, Section~\ref{section:sstequiv} will introduce a third scenario, which will be shown to be equivalent to the following two.

\subparagraph{I - Entanglement generation}
For encoding, Alice and Bob respectively prepare the states
$\ket{\Upsilon_1}^{AA'^n}$ and $\ket{\Upsilon_2}^{BB'^n},$
entangled with the $A'^n$ and $B'^n$ parts of the inputs of 
$\CN^{\otimes n}$.  Their goal is to do this in such a way 
so that Charlie, after applying a suitable decoding operation
$\CD\colon C^n\rightarrow \widehat{A}\widehat{B}$, can hold the $\widehat{A}\widehat{B}$ part of a state which is close to $\ket{\Phi_1}^{A\widehat{A}}\ket{\Phi_2}^{B\widehat{B}}$.   
Formally, $(\ket{\Upsilon_1}^{AA'^n}, \ket{\Upsilon_2}^{BB^n},\CD)$ is a $(2^{nR},2^{nS},n,\epsilon)$ \emph{qq entanglement generation code} for the channel $\CN$ if
\begin{eqnarray}
F(\ket{\Phi_1}\ket{\Phi_2},
\CD\circ\CN^{\otimes n}(\Upsilon_1\otimes\Upsilon_2)) \geq 1-\epsilon. \label{qqIs}
\end{eqnarray}
As before, $(R,S)$ is an \emph{achievable qq rate pair for entanglement generation} if there is a sequence of $(2^{nR},2^{nS},n,\epsilon_n)$ qq entanglement generation codes with $\epsilon_n\rightarrow 0$,  The capacity region $\CQ_\text{I}(\CN)$ is the closure of the collection of all such achievable rates.

\subparagraph{II - Entanglement transmission}
Alice and Bob each respectively have control over the $\tilde{A}$ and $\tilde{B}$ parts of the separate maximally entangled states
$\ket{\Phi_1}^{A\tilde{A}},\ket{\Phi_2}^{B\tilde{B}}$, while neither has access to $A$ or $B$.  Alice transfers the correlations in her system to the $A'^n$ parts of the inputs of $\CN^{\otimes n}$ with an encoding operation
$\CE_1\colon\tilde{A}\rightarrow A'^n$. Bob acts similarly with $\CE_2\colon \tilde{B}\rightarrow B'^n$.  Their goal is to preserve the respective correlations, 
so that Charlie can apply a decoding operation
$\CD\colon C^n\rightarrow \widehat{A}\widehat{B}$, in order to end up holding the $\widehat{A}\widehat{B}$ part of a state which is close to $\ket{\Phi_1}^{A\widehat{A}}\ket{\Phi_2}^{B\widehat{B}}$. 
Formally,
$(\CE_1,\CE_2,\CD)$ is a $(2^{nR},2^{nS},n,\epsilon)$ \emph{qq entanglement transmission code} for the channel $\CN$ if
\begin{eqnarray}
F(\ket{\Phi_1}\ket{\Phi_2},\CD\circ\CN^{\otimes n}\circ(\CE_1\otimes\CE_2)(\Phi_1\otimes \Phi_2))\geq 1-\epsilon. \label{qqIIs}
\end{eqnarray}
Achievable qq rate pairs for entanglement generation and the capacity region $\CQ_{\text{II}}(\CN)$ are defined as in the previous scenario.
As in the cq case, we defer to Section~\ref{section:sstequiv} the introduction of scenario III, as well as the proof that
\[\CQ(\CN) = \CQ_{\text{I}}(\CN) =  \CQ_{\text{II}}(\CN) =  \CQ_{\text{III}}(\CN).\]

\section{Main results} \label{section:mainresults}
Our first theorem gives a characterization of $\CC\CQ(\CN)$ as a regularized union of rectangles.

\begin{theo} \label{theo:cq}
Given a quantum multiple access channel $\CN\colon A'B'\rightarrow C$,
its cq capacity region $\CC\CQ(\CN)$ is given by the closure of
\[
\bigcup_{k=1}^\infty
\frac{1}{k}\CC\CQ^{(1)}(\CN^{\otimes k}),
\]
where $\CC\CQ^{(1)}(\CM)$
equals the pairs of nonnnegative rates $(r,S)$ satisfying
\begin{eqnarray*}
r &<& I(X;C)_\sigma \\
S &<& I_c(B\,\rangle CX)_\sigma
\end{eqnarray*}
for some pure state ensemble $\{p(x),\ket{\phi_x}^{A'}\}_{x\in\CX}$ and a bipartite pure state $\ket{\Psi}^{BB'}$ giving rise to
\begin{equation}
\sigma^{XBC} = \sum_x p(x) \proj{x}^X\otimes \CM(\phi_x\otimes\Psi). \label{th1arise}
\end{equation}
Furthermore, it is sufficient to consider $|\CX| \leq \min\{|A'|,|C|\}^2 + 1$ when computing $\CC\CQ^{(1)}$.
\end{theo}

The next theorem offers a characterization of $\CQ(\CN)$ as a regularized union of pentagons.
\begin{theo} \label{theo:qq}
Given a quantum multiple access channel $\CN\colon A'B'\rightarrow C$, its qq capacity region $\CQ(\CN)$ is given by the closure of
\[
\bigcup_{k=1}^\infty
\frac{1}{k}\CQ^{(1)}(\CN^{\otimes k}),
\]
where $\CQ^{(1)}(\CM)$ equals the
pairs of nonnegative rates $(R,S)$  satisfying
\begin{eqnarray*}
R &<& I_c(A \,\rangle BC)_\sigma\\
S &<& I_c(B\,\rangle AC)_\sigma \\
R+S &<& I_c(AB\,\rangle C)_\sigma
\end{eqnarray*}
for some bipartite pure states $\ket{\Psi_1}^{AA'}$ and $\ket{\Psi_2}^{BB'}$ giving rise to
\begin{equation}
\sigma^{ABC} = (1^{AB}\otimes\CM)(\Psi_1 \otimes\Psi_2). \label{th2arise}
\end{equation}

\end{theo}

\begin{figure}
 \centering
     \includegraphics[width=.4\textwidth]{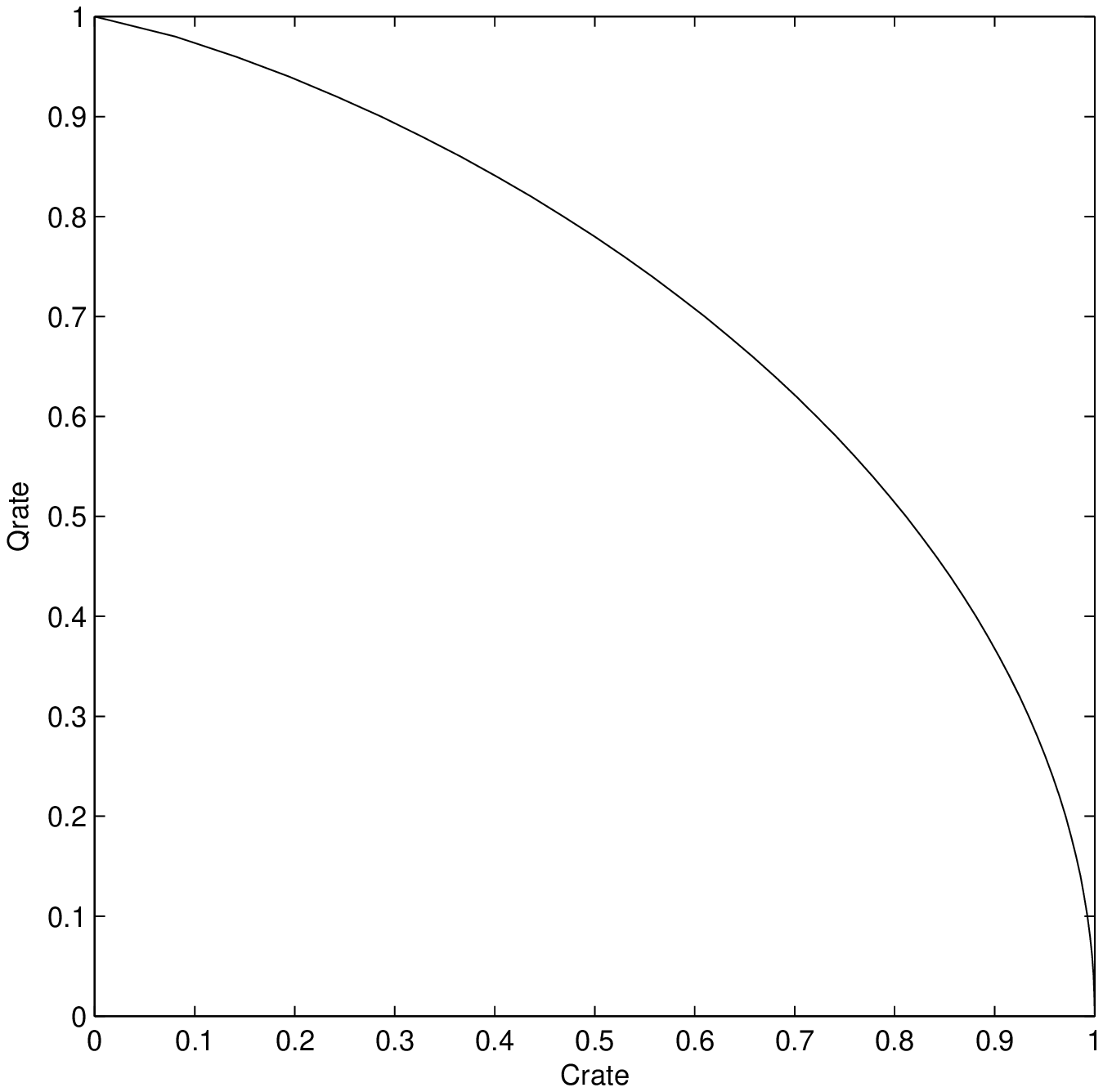}
     \caption{$\CC\CQ($erasure channel$)$}
     \label{erasure}
\end{figure}

We remark here that there does not appear to be any obstacle preventing application of the methods used in this paper to prove many-sender generalizations of the above theorems.  For simplicity, we have focused on the situations with two senders. It should also be noted that the characterizations given in each of the above theorems do not apparently lead to a finite computation for determining the capacity regions, as neither admits a single-letter characterization in general.  However, as an application, it is proved in Section~\ref{section:appendix:cqadd} that the cq capacity region for a certain quantum erasure multiple access channel
does in fact have a single-letter region, given by the set of all pairs of nonnegative classical-quantum rates $(r,S)$ satisfying
\begin{eqnarray*}
r &\leq& H(q) \\
S &\leq& (1-2q)\log d
\end{eqnarray*}
for some $0\leq q\leq \frac{1}{2}$.  This region is pictured in Figure~\ref{erasure} for the case in which $d=2$.  
In Section~\ref{section:appendix:qqadd}, we define a family of ``collective phase flip channels" which, with probability $p$, flip the phases of both input qubits.  We show there that the qq capacity region of such a channel is given by a single pentagon of nonnegative qq rates $(R,S)$ satisfying 
\begin{eqnarray*}
R &\leq& 1 \\ S &\leq& 1 \\ R+S &\leq& 2-H(p).   
\end{eqnarray*}

The characterizations given in Theorems~1 and 2 are not the only possible ways to describe the corresponding regions.  It is possible to prove coding theorems and converses for regularizations of distinct single-letter regions for each of $\CC\CQ(\CN)$ and $\CQ(\CN)$.  We refer the reader to Section~\ref{section:discussion} for further discussion regarding other characterizations of these regions.  
We finally mention that, contrary to the corresponding result for classical multiple access channels, the regions of Theorems 1 and 2 do not require convexification.  That this follows from the multi-letter nature of the regions will be demonstrated in the appendix.  

\section{Proofs of Theorems 1 and 2} \label{section:proofs}
We first collect some relevant results which will be used in what follows, starting with some relationships between our distance measures.  If $\rho$ and $\sigma$ are density matrices defined on the same (or isomorphic) Hilbert spaces, set
\[F = F(\rho,\sigma) \text{ and } T = |\rho - \sigma|_1.\]
Then, the following inequalities hold (see e.g.\ \cite{cn})
\begin{eqnarray}
1-\sqrt{F} &\leq \;\;T/2\;\; \leq&  \sqrt{1-F}, \label{trfid}\\
1-T &\leq \;\;F\;\; \leq& 1-T^2/4. \label{fidtr}
\end{eqnarray}
From these inequalities, we can derive the following more useful relationships
\begin{eqnarray}
F > 1-\epsilon &\Rightarrow& T \leq 2\sqrt{\epsilon} \label{trfide}\\
T \leq \epsilon &\Rightarrow& F > 1-\epsilon, \label{fidtre}
\end{eqnarray}
which are valid for $0\leq\epsilon\leq 1.$
Uhlmann \cite{uhl} has given the following characterization of fidelity
\[F(\rho,\sigma) = \max_{\ket{\Psi_\rho},\ket{\Phi_\sigma}} |\braket{\Psi_\rho}{\Phi_\sigma}|^2
= \max_{\ket{\Psi_\rho}}|\braket{\Psi_\rho}{\Phi_\sigma}|^2\]
where the first maximization is over all purifications of each state, and the second maximization holds for any fixed purification $\ket{\Phi_\sigma}$ of $\sigma.$
This characterization is useful in two different ways.  First, for any two states, it guarantees the existence of purifications of those states whose squared inner product equals the fidelity.  Second, one can derive from that characterization the following monotonicity property \cite{bcfjs} associated with an arbitrary trace-preserving channel $\CN$
\begin{eqnarray}
F(\rho,\sigma) &\leq& F(\CN(\rho),\CN(\sigma)). \label{fmon}
\end{eqnarray}
An analogous property is shared by the trace distance \cite{rustrace},
\begin{eqnarray}
|\rho-\sigma|_1 &\geq& \left|\CN(\rho)-\CN(\sigma)\right|_1, \label{trmon}
\end{eqnarray}
which holds even if $\CN$ is trace-reducing.  A simple proof for the trace-preserving case can be found in \cite{cn}.
These inequalities reflect the fact that completely-positive maps are \emph{contractive} and cannot improve the distinguishability of quantum states; the closer states are to each other, the harder it is to tell them apart.  We will often refer to either of these two properties as just ``montonicity," as the particular one to be used will always be clear from the context.  
Another useful property will be the multiplicativity of fidelities under tensor products
\begin{eqnarray}
F(\rho_1\otimes \rho_2,\sigma_1\otimes \sigma_2)
= F(\rho_1,\sigma_1)F(\rho_2,\sigma_2). \label{fmult}
\end{eqnarray}
Since the trace distance is a norm, it satisfies the triangle inequality.  The fidelity is not a norm, but it is possible to derive the following analog by applying (\ref{trfid}) and (\ref{fidtr}) to the triangle inequality for the trace distance
\begin{eqnarray}
F(\rho_1,\rho_3) \geq 1 - 2\sqrt{1-F(\rho_1,\rho_2)} - 2\sqrt{1-F(\rho_2,\rho_3)}.
\label{ftri}
\end{eqnarray}
It will be possible to obtain a sharper triangle-like inequality as a consequence of the following lemma, which states that if a measurement succeeds with high probability on a state, it will also do so on a state which is close to that state in trace distance.
\begin{lem}\label{lemma:special}
Suppose $\rho,\sigma,\Lambda \in \CB(\CH),$ where $\rho$ and $\sigma$ are density matrices, and $0\leq \Lambda \leq 1.$  Then, $\tr\Lambda\sigma \geq \tr\Lambda\rho - |\rho-\sigma|_1.$

\end{lem}
\emph{Proof:}
\begin{eqnarray*}
\tr\Lambda\sigma &=& \tr\Lambda\rho - \tr\Lambda(\rho - \sigma) \\
&\geq& \tr\Lambda\rho - \max_{0\leq\Lambda\leq 1} 2\tr\Lambda(\rho - \sigma)\\
&=& \tr\Lambda\rho - |\rho-\sigma|_1,
\end{eqnarray*}
where the last equality follows from a characterization of trace distance given in \cite{cn}.
\qed

Since $F(\phi,\rho) = \tr\phi\rho$ when $\phi$ is a pure state, a corollary of Lemma~\ref{lemma:special} is that
\begin{eqnarray*}
F(\phi,\sigma) \geq F(\phi,\rho) - |\rho - \sigma|_1,
\end{eqnarray*}
a fact we will refer to as the ``special triangle inequality."

The following lemma can be thought of either as a type of transitivity property inherent to any bipartite state with a component near a pure state, or as a partial converse to the monotonicity of fidelity.
\begin{lem}\label{lemma:transitivity}
For finite dimensional Hilbert spaces $\CH_A$ and $\CH_B$, let
$\ket{\phi}^A\in\CH_A$ be a pure state, $\rho^B\in\CB(\CH_B)$ a density matrix,
and $\Omega^{AB}\in\CB(\CH_A\otimes\CH_B)$ a density matrix with partial traces $\Omega^A = \tr_B\Omega$ and $\Omega^B = \tr_A\Omega,$ for which
\begin{eqnarray}
F(\phi,\Omega^A) \geq 1-\epsilon. \label{l4f1}
\end{eqnarray}
Then
\[F(\phi\otimes \rho, \Omega) \geq 1-|\rho - \Omega^B|_1 - 3\epsilon.\]
\end{lem}
\emph{Proof.}
We begin by defining the subnormalized density matrix $\tilde{\omega}$ via the equation
\begin{eqnarray}
(\phi\otimes 1)\Omega (\phi\otimes 1) = \phi\otimes \tilde{\omega}, \label{l4po}
\end{eqnarray}
which we interpret as the upper-left block of $\Omega$, when the basis for $\CH_A$ is chosen in such a way that $\ket{\phi} = (1,0,\ldots,0)^T.$  Notice that $F(\phi,\tr_B\Omega) = \tr\tilde{\omega}\equiv f.$
Writing the normalized state $\omega = \tilde{\omega}/f,$ we see that it
is close to $\tilde{\omega}$ in the sense that
\begin{eqnarray}
|\omega - \tilde{\omega}|_1 &\leq& \epsilon|\tilde{\omega}|_1
  \nonumber \\
&\leq& \epsilon. \label{l4oo}
\end{eqnarray}
Now we write
\begin{eqnarray}
\sqrt{F(\phi\otimes\rho,\Omega)}
&=& \tr\sqrt{\sqrt{(\phi\otimes\rho)}\Omega\sqrt{(\phi\otimes\rho)}}
  \nonumber\\
&=& \tr\sqrt{(1\otimes\sqrt{\rho})
  (\phi\otimes 1)\Omega(\phi\otimes 1)
  (1\otimes\sqrt{\rho})}\nonumber\\
&=& \tr\sqrt{(1\otimes\sqrt{\rho})(\phi\otimes\tilde{\omega})(1\otimes\sqrt{\rho})}
  \nonumber\\
&=& \tr\sqrt{\phi\otimes(\sqrt{\rho}\,\tilde{\omega}\sqrt{\rho})}\nonumber\\
&=& \tr\sqrt{\sqrt{\rho}\,\tilde{\omega}\sqrt{\rho}}\nonumber\\
&=& \sqrt{F(\tilde{\omega},\rho)}\nonumber\\
&=& \sqrt{f F(\omega,\rho)}\nonumber\\
&\geq& \sqrt{f (1-|\omega - \rho|_1)}. \label{l4F}
\end{eqnarray}
The first line is the definition of fidelity and the third follows from (\ref{l4po}).
The last equality relies on the fact that the fidelity, as we've defined it, is linear in either of its two inputs,
while the inequality follows from (\ref{fidtr}).

Noting that $\Omega^B \geq \tilde{\omega}$, we define another positive operator
$\omega' = \Omega^B - \tilde{\omega},$ which by (\ref{l4f1}) satisfies $\tr\omega' \leq \epsilon$ and can be interpreted as the sum of the rest of the diagonal blocks of $\Omega.$
The trace distance in the last line above can be bounded via double application of the triangle inequality as
\begin{eqnarray}
|\rho - \omega|_1
&\leq& |\rho - (\rho-\omega')|_1 + |(\rho-\omega') - \tilde{\omega}|_1 +
   |\tilde{\omega} -\omega|_1 \nonumber \\
&\leq& \tr\omega' + \left|\rho - \Omega^B\right|_1 + \epsilon \nonumber\\
&\leq& \left|\rho - \Omega^B\right|_1 + 2\epsilon \label{l4ro},
\end{eqnarray}
where the second line follows from (\ref{l4oo}).
Combining (\ref{l4F}) with (\ref{l4ro}), we obtain
\begin{eqnarray*}
F(\phi\otimes\rho,\Omega)
&\geq& (1-\epsilon)(1-|\rho - \Omega^B|_1-2\epsilon)\\
&\geq& 1-\left|\rho - \Omega^B\right|_1-3\epsilon.
\end{eqnarray*}
\qed

This continuity lemma from \cite{alickifannes} shows that if two bipartite states are close to each other, the difference between their associated coherent informations is small.
\begin{lem}[Continuity of coherent information]\label{lemma:continuity}
Let $\rho^{QR}$ and $\sigma^{QR}$ be two states of a
finite-dimensional bipartite system $QR$ satisfying
$|\rho-\sigma|_1 \leq \epsilon$. Then
\[|I_c(Q\,\rangle R)_\rho - I_c(Q\,\rangle R)_\sigma|
   \leq 2H(\epsilon) + 4\log|Q|\epsilon,\]
where $H(\epsilon)$ is the binary entropy function.
\end{lem}

Next is Winter's ``gentle measurement" lemma \cite{gentle}, which implies that a measurement which is likely to be successful in identifying a state tends not to significantly disturb that state.
\begin{lem}[Gentle measurement]\label{lemma:gentle}
Let $\CH$ be a finite dimensional Hilbert space. If $\rho\in\CB(\CH)$ is a density matrix and $\Lambda\in\CB(\CH)$ is nonnegative with spectrum bounded above by 1,  then
\[\tr \rho\Lambda \geq 1-\epsilon\]
implies
\[\left|\sqrt{\Lambda}\rho\sqrt{\Lambda} - \rho\right|_1 \leq \sqrt{8\epsilon}.\]
\end{lem}

Our coding theorems for multiple access channels will make use of existing coding theorems for single-user channels.
$(\CE,\CD)$ is a $(2^{nR},n,\epsilon)$ \emph{entanglement transmission code} for the channel $\CN$ if, for the $2^{nR}\times 2^{nR}$ maximally entangled state $\ket{\Phi}^{A\tilde{A}}$, we have
\[F(\ket{\Phi},\CD\circ\CN^{\otimes n}\circ\CE(\Phi)) \geq 1-\epsilon.\]
A $(2^{nR},n,\epsilon)$ \emph{random entanglement transmission code} consists of an collection of deterministic $(2^{nR},n,\epsilon)$ entanglement transmission codes
$(\CE_\beta,\CD_\beta)$ and a probability distribution $P_\beta$, corresponding to a source of shared common randomness available to both sender and receiver.  We will often omit the subscript, once the randomness of the code has been clarified, and it will be understood that $\CE$ and $\CD$ constitute a pair of correlated random maps.

Associated to a random code is its expected, or average code density operator $\E\CE(\pi^{\tilde{A}})$, which is the expectation, over the shared randomness, of the image of the maximally mixed state on $\tilde{A}$.  Our reason for using random quantum codes will be to ensure that, on average, the input to $\CN^{\otimes n}$ is at least close to a $n$-fold product state.

The proof of the existence of quantum codes achieving the coherent information bound is attributed to Lloyd \cite{lloyd}, Shor \cite{shor} and Devetak \cite{dev}.
The following quantum coding proposition for single-user channels is proved in \cite{dev} and concerns the existence of random entanglement transmission codes whose average code density matrix can be made arbitrarily close to a product state. 
\begin{prop} \label{prop:lsd}
Given is a channel $\CN\colon A'\rightarrow B$, a density matrix $\rho^{A'}$, and a number $0\leq R < I_c(\rho,\CN).$  For every $\epsilon>0$, there is $n$ sufficiently large so that there is a $(2^{nR},n,\epsilon)$ random entanglement transmission code $(\CE,\CD)$ for $\CN$
with an isometric encoder $\CE$ and average code density operator $\varrho^{A'^n} = \E\CE(\pi^A)$ satisfying
\[|\varrho - \rho^{\otimes n}|_1\leq\epsilon.\]
Furthermore, given any particular isometric extension 
$\CU_\CN \colon A'\rightarrow BE$ of $\CN$, it is possible to choose isometric extensions $\CU_{\CD}^\beta\colon B^n\rightarrow \widehat{A}F$ of the deterministic decoders so that 
\begin{equation}
F\big(\ket{\Phi}^{A\widehat{A}}\ket{\lambda}^{E^nF},\CU_{\CD}^\beta\circ
\CU_\CN^{\otimes n}\circ\CE^\beta\ket{\Phi}^{A\tilde{A}}\big)\geq 1-\epsilon
\label{prop1fid}
\end{equation} 
for some fixed pure state $\ket{\lambda}^{E^nF}$. 
\end{prop}

Next, we state an average error version of the HSW Theorem for cq codes with codewords chosen i.i.d.\ according to a product distribution \cite{sw2,holcap}.
\begin{prop}[HSW Theorem] \label{prop:hsw}
Given is a cq state
$\sigma^{XQ}=\sum_x p(x)\proj{x}^X\otimes\rho^Q_x$ and a number $0\leq R < I(X;Q)_\sigma.$ For every $\epsilon > 0$, there
is $n$ sufficiently large so that if $2^{nR}$ codewords $\CC=\{X^n(m)\}$ are chosen i.i.d.\ according to the product distribution $p(x^n) = \prod_{i=1}^n p(x_i)$, corresponding to input preparations
\[\rho_{x^n} = \rho_{x_1}\otimes\cdots\otimes\rho_{x_n},\] there exists a decoding POVM $\{\Lambda_m\}$ on $Q^n$, depending on the random choice of codebook $\CC$, which correctly identifies the index $m$ with average probability of error less than $\epsilon,$ in the sense that
\begin{eqnarray}
\E_\CC 2^{-nR}\sum_{m=1}^{2^{nR}} \tr\rho_{X^n(m)} \Lambda_m
\geq 1-\epsilon.
\end{eqnarray}
\end{prop}
Due to the symmetry of the distribution of $\CC$ under codeword permutations, it is clear that the expectations of each term in the above sum are equal.  In other words,
\begin{eqnarray}
\E_\CC 2^{-nR}\sum_{m=1}^{2^{nR}} \tr\rho_{X^n(m)} \Lambda_m
= \E_\CC \tr\rho_{X^n(1)} \Lambda_1, \label{symmetry}
\end{eqnarray}
so we will later, without loss of generality, make the assumption that Alice sends codeword $M=1$ during our analysis (see \cite{coverthomas} for a detailed discussion in the classical case).

\paragraph{Proof of Theorem~\ref{theo:cq} (converse)}
We prove in Section~\ref{section:sstequiv} that any rate pair which is achievable for entanglement transmission is also achievable for entanglement generation.  For this reason, we use the latter scenario to prove the converse part of Theorem~\ref{theo:cq}.
It should be noted that the reverse implication, namely that entanglement generation implies entanglement transmission, follows from the fact the outer bound to be proved next coincides with the inner bound obtained by the coding theorem below.

Suppose there exists a sequence of $(2^{nr},2^{nS},n,\epsilon_n)$
entanglement generation codes with $\epsilon_n\rightarrow 0$.  Fixing a blocklength $n$, let $\{\ket{\phi_m}\}_{m\in 2^{nr}}$, $\ket{\Upsilon}^{BB'^n}$ and $\boldsymbol{\CD}$ comprise the corresponding cq entanglement generation code.
The state induced by the encoding is
\[\omega^{MBC^n} = 2^{-nr}\sum_{m=1}^{2^{nr}}
\proj{m}^M\otimes (1^B\otimes\CN^{\otimes n})(\phi_m\otimes \Upsilon).\]
After application of the decoding instrument $\boldsymbol{\CD}\colon C^n\rightarrow \widehat{B}\widehat{M}$, this state becomes
\[\Omega^{M\widehat{M}B\widehat{B}} = (1^{MB}\otimes\boldsymbol{\CD})(\omega).\]
An upper bound on the classical rate of the code can be obtained as follows:
\begin{eqnarray*}
nr &=& H(M)_\Omega \\
&=& I(M;\widehat{M})_\Omega + H(M|\widehat{M})_\Omega \\
&\leq& I(M;\widehat{M})_\Omega + H(\epsilon_n) + nr\epsilon_n \\
&\leq& I(M;C^n)_\omega + n\epsilon'_n.
\end{eqnarray*}
The first inequality follows from Fano's inequality (see e.g. \cite{coverthomas}) while in the second we use the Holevo bound \cite{holevobound} and define $\epsilon_n' = \frac{1}{n} + r\epsilon_n$.
The quantum rate of the code is upper bounded as
\begin{eqnarray*}
I_c(B\,\rangle C^nM)_\omega &\geq& I_c(B\,\rangle \widehat{B}M)_\Omega\\
&\geq& I_c(B\,\rangle \widehat{B})_\Omega \\
&\geq& I_c(B\,\rangle \widehat{B})_\Phi - 2H(\epsilon_n) - 8nS\sqrt{\epsilon_n} \\
&=& nS - n\epsilon_n''.
\end{eqnarray*}
Above, the inequalities are consequences of the data processing inequality \cite{sn96}, the fact that conditioning cannot increase entropy (and thus  cannot decrease coherent information) \cite{cn}, a combination of Lemma~\ref{lemma:continuity} and (\ref{trfide}), and the definition
$\epsilon_n'' = \frac{2}{n} + nS\sqrt{\epsilon_n}$.
The second justification can be considered as an alternative statement of the well-known strong subadditivity inequality \cite{ssuborig}, of which a recent simple proof can be found in \cite{ssub}. 
Setting $X=M$, we have thus proven that
\[r \leq \frac{1}{n}I(X;C^n) + \epsilon'_n, \,\,\,\,\,\,
S\leq \frac{1}{n}I_c(B\,\rangle C^nX) + \epsilon''_n\]
whenever $(r,S)$ is an achievable cq rate pair for entanglement generation, where $\epsilon'_n,\epsilon''_n\rightarrow 0$. 
It follows that for any achievable rate pair $(r,S)$ and any $\delta>0$, we have \[(r-\delta,S-\delta) \in 
\frac{1}{n}\CC\CQ^{(1)}(\CN^{\otimes n}) \subseteq \CC\CQ(\CN).\]
Since $\CC\CQ(\CN)$ is closed by definition, this completes the proof.

\qed

\paragraph{Proof of Theorem~\ref{theo:cq} (achievability)}
Our method of proof for the coding theorem will work as follows.  We will employ random HSW codes and random entanglement transmission codes to ensure that the average state at the input of $\CN^{\otimes n}$ is close to a product state.  Each sender will utilize a code designed for the product channel induced by the other's random input, whereby
existing coding theorems for product channels will be invoked.
The quantum code used will be one which achieves the capacity of a modified channel, in which the classical input is copied, without error, to the output of the channel.
As the random HSW codes will exactly induce a product state input, the existence of these quantum codes will follow directly from Proposition \ref{prop:lsd}.

The random HSW codes will be those which exist for product channels.  As random entanglement transmission codes exist with average code density matrix arbitrarily close to a product state, this will ensure that the resulting output states are distinguishable with high probability.  Furthermore, obtaining the classical information will be shown to cause but a small disturbance in the overall joint quantum state of the system.  As we will show, it is possible to mimic the channel for which the quantum code is designed by placing the identities of the estimated classical message states into registers appended to the outputs of each channel in the product.

The decoder for the modified channel will then be shown to define a quantum instrument which satisfies the success condition for a cq entanglement transmission code, on average.  This feature will then be used to
infer the existence of a particular, deterministic code which meets the same requirement.

Fix a pure state ensemble $\{p(x),\ket{\phi_x}^{A'}\}$ and a bipartite pure state $\ket{\Psi}^{B''B'}$ which give rise to the cqq state
\[\omega^{XB''C} =
\sum_x p(x)\proj{x}^X\otimes(1^B\otimes \CN)
(\phi_x^{A'}\otimes \Psi^{B''B'}),\]
which has the form of (\ref{th1arise}).
Define $\rho_1^{A'} = \sum_x p(x)\phi_x$ and $\rho_2^{B'} = \tr_B\Psi$.   We will demonstrate the achievability of the corner point $(I(X;C),I_c(B''\,\rangle CX))_\omega$ by showing that for every $\epsilon,\delta>0,$ if $r = I(X;C)_\omega-\delta$ and $S = I_c(B''\,\rangle CX)_\omega-\delta$,
there exists a $(2^{nr},2^{nS},n,\epsilon)$ cq entanglement transmission code for the channel $\CN$, provided that $n$ is sufficiently large and that $S> 0$.
The rest of the region will follow by timesharing.

For encoding, Alice will choose $2^{nr}$ sequences $X^n(m)$, i.i.d.\ according to the product distribution $p(x^n) = \prod_{i=1}^n p(x_i)$.  As each sequence corresponds to a preparation of channel inputs
$\ket{\phi_m}^{A'^n}=
\ket{\phi_{X_1(m)}}\otimes\cdots\otimes\ket{\phi_{X_n(m)}},$ the expected  average density operator associated with Alice's input to the channel is precisely
\[
\E_\CC 2^{-nr}\sum_{m=1}^{2^{nr}}\proj{\phi_m} = \sum_{x^n} p(x^n) \proj{\phi_{x^n}}
=\rho_1^{\otimes n}.
\]
Define a new channel $\boldsymbol{\CN}\!_2\colon B'\rightarrow C\widehat{X}$ (which is also an instrument) by
\[\boldsymbol{\CN}\!_2\colon \rho\mapsto
\sum_x p(x)\CN(\phi_x\otimes\rho)\otimes\proj{x}^{\widehat{X}}.\]
This can be interpreted as a channel which reveals the identity of Alice's input state to Charlie, with the added assumption that Alice chooses her inputs at random.  Alternatively, one can view this as a channel with state information available to the receiver, where nature is randomly choosing the ``state" $x$ at Alice's input.  Observe that 
$I_c(\rho_2,\boldsymbol{\CN}\!_2) = I_c(B''\,\rangle CX)$.
By Proposition \ref{prop:lsd}, there exists a $(2^{nS},n,\epsilon)$ random entanglement transmission code $\{\CE,\CD,\beta\}$ for the channel $\boldsymbol{\CN}\!_2$, with average code density operator
$\varrho^{B'^n} = \E_\beta\CE(\pi)$ satisfying \[|\varrho - \rho_2^{\otimes n}|_1\leq \epsilon.\]

Now, by Proposition \ref{prop:hsw}, for the channel $\CN_1\colon \rho\mapsto \CN(\rho\otimes\rho_2)$ which would result if Bob's average code density operator were \emph{exactly} equal to $\rho_2^{\otimes n},$ there exists a decoding POVM $\{\Lambda_m\}_{m\in 2^{nr}}$ which would identify Alice's index $m$ with expected average probability of error less than $\epsilon$,
in the sense that
\[\E_\CC 2^{-nr}\sum_{m=1}^{2^{nr}}\tr\Lambda_m\tau'_m \geq 1-\epsilon,\]
where
\[\tau'_m = \CN^{\otimes n}(\phi_m\otimes\rho_2^{\otimes n}).\]
By the symmetry of the random code construction, we utilize (\ref{symmetry}) to write this as
\[\E_\CC\tr\Lambda_1\tau'_1 \geq 1-\epsilon.\]
Define the \emph{actual} output of the channel corresponding to $M=m$ as
\[\tau_m = \CN^{\otimes n}(\phi_m\otimes\CE(\pi)),\]
as well as its extension
\[\xi_m^{BC^n} = \CN^{\otimes n}(\phi_m\otimes\CE(\Phi)),\]
where $\ket{\Phi}^{B\tilde{B}}$ is the maximally entangled state which Bob is required to transmit.
Note that
\[\E_\beta\tau_m = \E_\beta\tr_B\xi_m = \CN^{\otimes n}(\phi_m\otimes\varrho).\]
It follows from monotonicity of trace distance that
\[\left|\E_\beta\tau_1 - \tau_1'\right|_1\leq \epsilon,\]
which, together with Lemma~\ref{lemma:special}, implies that
\begin{eqnarray*}
\E_\CC 2^{-nr}\sum_{m=1}^{2^{nr}}\tr\Lambda_m \E_\beta\tau_m
= \E_{\CC\beta}\tr\Lambda_1 \tau_1
\geq 1-2\epsilon.
\end{eqnarray*}
This allows us to bound the expected probability of correctly decoding Alice's message as
\begin{eqnarray}
\E_{\CC\beta}\tr(1\otimes\Lambda_1)\xi_1 \geq 1-2\epsilon. \label{EP}
\end{eqnarray}

In order to decode, Charlie begins by performing the measurement $\{\Lambda_m\}_{m\in 2^{nr}}.$  He declares Alice's message to be $\widehat{M} = m$ if measurement result $m$ is obtained.  Charlie will then attempt to simulate the channel $\boldsymbol{\CN}\!_2^{\,\,\!\otimes n}$, by associating a separate classical register $\widehat{X}_i$ to each channel $\CN\colon A'_i\rightarrow C_i$ in the product, preparing the states $\ket{X_i(m)}^{\widehat{X}_i}$, for each $1\leq i \leq n$.
Additionally, he stores the result of the measurement in the system $\widehat{M}$, his declaration of the message intended by Alice.
This procedure results in the global state
\[\Gamma^{BC^n\widehat{X}^n\widehat{M}} =
\sum_{m=1}^{2^{nr}}\left(1\otimes\sqrt{\Lambda_m}\right)\xi_1\left(1\otimes\sqrt{\Lambda_m}\right)
\otimes \proj{X^n(m)}^{\widehat{X}^n}\otimes \proj{m}^{\widehat{M}}.\]
Let $\Upsilon^{BC^n\widehat{X}^n} = \tr_{\widehat{M}} \Gamma$.
If Charlie were able to perfectly reconstruct Alice's classical message, $\Gamma$ would instead be
\[\Gamma' = \xi_1 \otimes \proj{X^n(1)}^{\widehat{X}^n}\otimes\proj{1}^{\widehat{M}},\]
with $\Upsilon' = \tr_{\widehat{M}}\Gamma'$.
When averaged over Alice's random choice of HSW code,
$\Upsilon'$ is precisely equal to the state which would arise via the action of the modified channel $\boldsymbol{\CN}\!_2.$ This is because
\begin{eqnarray}
\E_\CC \Upsilon' &=&
\sum_{x^n} p(x^n) \xi_{x^n}\otimes\proj{x^n}^{\widehat{X}^n} \nonumber \\
&=& \boldsymbol{\CN}\!_2^{\,\,\!\otimes n}\circ\CE(\Phi), \label{ECU}
\end{eqnarray}
where we have written the state which results when Alice prepares  $\phi_{x^n}$ as
\[\xi_{x^n} = \CN^{\otimes n}(\phi_{x^n}\otimes \CE(\Phi)).\]
However, our choice of a good HSW code ensures that he can almost perfectly reconstruct Alice's message.  A consequence of this will be that the two states $\Upsilon$ and $\Upsilon'$ are almost the same, as we will now demonstrate.

In what follows, we will need to explicitly keep track of the randomness
in our codes, by means of superscripts which are to be interpreted as indexing the deterministic codes which occur with the probabilities $P_\CC$ and $Q_\beta$.
Rewriting (\ref{EP}) as
\[\sum_{\CC\beta} P_\CC Q_\beta
 \tr\left(1\otimes\Lambda_1^\CC\right)\xi_1^{\CC\beta}\geq 1-2\epsilon,\]
it is clear that we may write
\[\tr\left(1\otimes\Lambda_1^\CC\right)\xi_1^{\CC\beta} \geq  1-\epsilon_{\CC\beta},\]
for positive numbers $\{\epsilon_{\CC\beta}\}$ chosen to satisfy
\[\sum_{\CC\beta} P_\CC Q_\beta \epsilon_{\CC\beta} = 2\epsilon.\]
By the gentle measurement lemma,
\[\left| \left(1\otimes\sqrt{\Lambda_1^\CC}\right)
    \xi_1^{\CC\beta}
   \left(1\otimes\sqrt{\Lambda_1^\CC}\right)
   - \xi_1^{\CC\beta}\right|_1 \leq \sqrt{8\epsilon_{\CC\beta}},\]
and thus, by the concavity of the square root function,
\begin{eqnarray*}
\E_{\CC\beta} \left|\left(1\otimes\sqrt{\Lambda_1}\right) \xi_1 \left(1\otimes\sqrt{\Lambda_1}\right) - \xi_1\right|_1
 &=& \sum_{\CC\beta} P_\CC Q_\beta
\left|\left(1\otimes\sqrt{\Lambda_1^\CC}\right)
    \xi_1^{\CC\beta}
   \left(1\otimes\sqrt{\Lambda_1^\CC}\right)
   - \xi_1^{\CC\beta}\right|_1 \\
 &\leq& 4\sqrt{\epsilon}.
\end{eqnarray*}
Along with (\ref{EP}) and monotonicity with respect to $\tr_{\widehat{M}}$, this estimate allows us to express
\begin{eqnarray}
\E_{\CC\beta}|\Upsilon - \Upsilon'|_1 
&\leq& \E_{\CC\beta}|\Gamma - \Gamma'|_1 \\
&=& \E_{\CC\beta}\left|\left(1\otimes\sqrt{\Lambda_1}\right)
      \xi_1\left(1\otimes\sqrt{\Lambda_1}\right) - \xi_1\right|_1 
      \nonumber \\
& &  + \E_{\CC\beta}\sum_{m=2}^{2^{nr}}
\left|\left(1\otimes\sqrt{\Lambda_m}\right)
      \xi_1\left(1\otimes\sqrt{\Lambda_m}\right)\right|_1  \nonumber \\
&=&\E_{\CC\beta}\left|\left(1\otimes\sqrt{\Lambda_1}\right)
      \xi_1\left(1\otimes\sqrt{\Lambda_1}\right) - \xi_1\right|_1
+ \E_{\CC\beta}\sum_{m=2}^{2^{nr}} \tr(1\otimes\Lambda_m)\xi_1  \nonumber \\
&\leq& 4\sqrt{\epsilon} + 2\epsilon  \nonumber \\
&\leq& 5\sqrt{\epsilon}, \label{ECbU}
\end{eqnarray}
provided that $\epsilon\leq \frac{1}{2}.$
Since the the entanglement fidelity is linear in $\CD(\Upsilon),$ which is itself linear in $\Upsilon,$ we can also use the special triangle inequality to write
\begin{eqnarray*}
F(\ket{\Phi},\CD(\E_{\CC\beta}\Upsilon))
&=& F(\ket{\Phi},\E_\beta\CD(\E_\CC\Upsilon))\\
&\geq& F\big(\ket{\Phi},\E_\beta\CD(\E_\CC\Upsilon')\big)
-\big|\E_\beta\CD(\E_\CC\Upsilon')-\E_\beta\CD(\E_\CC\Upsilon)\big|_1.
\end{eqnarray*}
Using our earlier observation from (\ref{ECU}) and the definition of a $(2^{nS},n,\epsilon)$ entanglement transmission code, we can bound the first term as
\begin{eqnarray*}
F(\ket{\Phi},\CD(\E_\CC\Upsilon'))
&=& F(\ket{\Phi},\CD\circ\boldsymbol{\CN}\!_2^{\,\,\!\otimes n}\circ\CE(\Phi)) \\
&\geq& 1-\epsilon.
\end{eqnarray*}
An estimate on the second term is obtained via
\begin{eqnarray*}
\left|\E_\beta\CD(\E_\CC\Upsilon) - \E_\beta\CD(\E_\CC\Upsilon') \right|_1
&\leq& \E_\beta\left|\CD(\E_\CC\Upsilon)-\CD(\E_\CC\Upsilon') \right|_1\\
&\leq& \E_\beta\left|\E_\CC\Upsilon - \E_\CC\Upsilon'\right|_1\\
&\leq& \E_{\CC\beta}\left|\Upsilon - \Upsilon'\right|_1\\
&\leq& 5\sqrt{\epsilon},
\end{eqnarray*}
where first three lines are by convexity, monotonicity, and convexity once again of the trace norm.  The last inequality follows from (\ref{ECbU}).
Putting these together gives
\begin{eqnarray}
\E_{\CC\beta}F(\ket{\Phi},\CD(\Upsilon))
&\geq& 1-\epsilon - 5\sqrt{\epsilon} \nonumber\\
&\geq& 1- 6\sqrt{\epsilon}. \label{EF}
\end{eqnarray}
At last, observe that the final decoded state $\Omega$ (which still depends on both sources of randomness $\CC$ and $\beta$) is equal to 
\[\Omega^{B\widehat{B}\widehat{M}} = 
\CD(\Gamma^{BC^n\widehat{X}^n\widehat{M}}) 
\equiv \boldsymbol{\CD}(\xi_1^{BC^n}),\]
implicitly defining the desired decoding instrument $\boldsymbol{\CD}\colon C^n \rightarrow \widehat{B}\widehat{M}$.
The expectation of (\ref{cqIIs}) can now be bounded as   
\begin{eqnarray*}
\E_{\CC\beta} 2^{-nr}\sum_{m=1}^{2^{nr}} P_s^{\text{II}}(m) &=&
\E_{\CC\beta} P_s^{\text{II}}(1) \\
&=& F(\ket{1}\ket{\Phi}, \E_{\CC\beta}\Omega) \\
&\geq& 
1 - \left|\tr_{B\widehat{B}} \E_{\CC\beta}\Gamma - \proj{1}\right|_1
- 3\big(1 - F(\ket\Phi,\CD(\Upsilon))\big) \\
&\geq& 1 - 2\sqrt{2\epsilon} - 18\sqrt{\epsilon} \\
&\geq& 1 - 21\sqrt{\epsilon}.
\end{eqnarray*}
The third line above is by Lemma \ref{lemma:transitivity}. The first estimate in the fourth line follows from (\ref{EP}), while the second estimate is by (\ref{EF}), together with (\ref{trfide}).
We may now conclude that there are particular values of the randomness indices $\beta$ and $\CC$ such that the same bound is satisfied for a deterministic code.
We have thus proven that $(\{\ket{\phi_m}\}_{m\in 2^{nr}},\CE,\boldsymbol{\CD})$ comprises a
$(2^{nr},2^{nS},n,21\sqrt{\epsilon})$ entanglement transmission code.  This concludes the coding theorem.

\qed

\paragraph{Proof of Theorem~\ref{theo:qq} (converse)}
Suppose that $(R,S)$ is an achievable qq rate pair for entanglement generation.  By definition, this means that there must exist a sequence of $(2^{nR},2^{nS},n,\epsilon_n)$ entanglement generation codes with $\epsilon_n\rightarrow 0$.  Fixing a blocklength $n$,
let $\ket{\Upsilon_1}^{AA'^n}, \ket{\Upsilon_2}^{BB'^n}$ and
$\CD\colon C^n\rightarrow \widehat{A}\widehat{B}$ comprise the corresponding encodings and decodings. Define
\[\omega^{ABC^n} =
(1^{AB}\otimes\CN^{\otimes n})(\Upsilon_1\otimes\Upsilon_2)\]
to be the result of sending the respective $A'^n$ and $B'^n$ parts of $\ket{\Upsilon_1}$ and $\ket{\Upsilon_2}$ through the channel $\CN^{\otimes n}$.  Further defining
\[\Omega^{AB\widehat{A}\widehat{B}} = (1^{AB}\otimes\CD)(\omega)\]
as the corresponding state after decoding,
the entanglement fidelity of the code is given by
\begin{eqnarray}
F_{AB}=F(\ket{\Phi_1}\otimes\ket{\Phi_2},\Omega) \geq 1-\epsilon_n. 
\label{fabconv}
\end{eqnarray}
where $\ket{\Phi_1}^{A\widehat{A}}$ and $\ket{\Phi_2}^{B\widehat{B}}$
are the maximally entangled target states.  The sum rate can be bounded as 
\begin{eqnarray*}
I_c(AB\,\rangle C^n)_\om 
&\geq& I_c(AB\,\rangle \h{A}\h{B})_\Om \\
&\geq& I_c(AB\,\rangle \h{A}\h{B})_{\Phi_1\!\otimes\Phi_2} 
- 2H(\ep_n) - 8n(R+S)\sqrt{\ep_n}\\ 
&\geq& n(R+S) - n\ep'_n.
\end{eqnarray*}
The first step is by the data processing inequality.  
The second step uses 
Lemma~\ref{lemma:continuity} and (\ref{trfide}), along with monotonicity applied to (\ref{fabconv}).  The last step has defined 
$\ep'_n = \frac{2}{n} - 8(R+S)\sqrt{\ep_n}$ and holds because the binary entropy $H(\cdot)$ is upper bounded by 1. 
We can bound Alice's rate $R$ by writing
\begin{eqnarray*}
I_c(A\,\rangle BC^n)_\omega
&\geq& 
I_c(A\,\rangle C^n)_\omega \\
&\geq& 
  I_c(A\,\rangle \widehat{A}\widehat{B})_{\Omega} \\
&{\geq}& 
  I_c(A \,\rangle \widehat{A})_{\Omega} \\
&{\geq}& 
  I_c(A \,\rangle \widehat{A})_{\Phi_1} -2H(\epsilon_n) - 8nR\sqrt{\ep_n} \\
 &\geq&  nR - n\ep'_n.
\end{eqnarray*}
The first three steps above are by data processing \cite{sn96}.  The remaining steps hold for the same reasons as in the previous chain of inequalities.  Similarly, Bob's rate also must satisfy 
\[nS\leq I_c(B\,\rangle AC^n)_\om + n\ep'_n. \]
Since $\epsilon_n\rightarrow 0$ implies $\epsilon'_n\rightarrow 0,$
this means that for every $\delta > 0$, any achievable qq rate pair $(R,S)$ must satisfy 
\[(R-\delta,S-\delta) \in \frac{1}{n}\CQ^{(1)}(\CN^{\otimes n}) \subseteq \CQ(\CN).\]
Since $\CQ(\CN)$ is closed by definition, this completes the proof.

\qed

\paragraph{Remark} Strictly speaking, the pair of nonnegative rates $(R,S)$ needs to be contained in some pentagon whose corner points 
$\big(\frac{1}{k}I_c(A\,\rangle C^k)_\sig,\frac{1}{k}I_c(B\,\rangle AC^k)_\sig\big)$ and 
$\big(\frac{1}{k}I_c(A\,\rangle BC^k)_\sig,\frac{1}{k}I_c(B\,\rangle C^k)_\sig\big)$ are located in the upper right quadrant of $\mathbb{R}^2$, where $\sigma^{ABC^k}$ is some state of the form (\ref{th2arise}).  For large enough $n$, the induced states $\omega$ in the above proof fulfill this role.  To see this, note that an artifact of the steps which upper bound Alice's rate $R$ is that $\frac{1}{n}I_c(A\,\rangle BC^n)\geq R-\ep_n$ and $I_c(A\,\rangle C^n) \geq R - \epsilon_n.$  Since $\ep_n\rightarrow 0$, the right sides are eventually positive whenever $R>0$.  The similar steps which bound Bob's rate complete the argument.

\paragraph{Proof of Theorem~\ref{theo:qq} (achievability)}
Fix bipartite pure states $\ket{\Psi_1}^{A''A'}$ and $\ket{\Psi_2}^{B''B'}$ which give rise to the state
\[\omega^{A''B''C} = (1^{A''B''}\otimes\CN^{\otimes n})(\Psi_1\otimes\Psi_2),\]
and define $\rho_1^{A'} = \tr_A\Psi_1$, $\rho_2^{B'} = \tr_B\Psi_2.$
Letting $\epsilon, \delta>0$ be arbitrary,
we will show that there exists a $(2^{nR},2^{nS},n,\epsilon)$ qq entanglement transmission code where
\[R=I_c(A''\,\rangle C)_\omega-\delta \text{ and }
S=I_c(B''\,\rangle A''C)_\omega-\delta\]
provided that $R,S \geq 0$.
Note that the rates in Theorem~\ref{theo:qq} will be implied by taking the channel to be $\CN^{\otimes k},$ with $\omega^{ABC^k}$ defined similarly.

Let us begin by choosing an isometric extension $\CU_\CN:A'B'\rightarrow CE$ of $\CN$.  
Define the ideal channel $\CN_1\colon A'\rightarrow C$ which would effectively be seen by Alice were Bob's average code density operator exactly equal to $\rho_2^{\otimes n}$  as
\[\CN_1\colon \tau\mapsto \CN(\tau \otimes\rho_2).\]
We now use $\CU_\CN$ to define a particular isometric extension 
$\CU_{\CN_1}\colon A'\rightarrow CE'$ of $\CN_1$, where $E'=B''E$, as 
\[\CU_{\CN_1}\colon\tau \mapsto \CU_\CN(\tau\otimes \Psi_2).\]
Observe that Bob's fake input $B''$ is treated as part of the environment of Alice's ideal induced channel.   
We then further define the channel $\CN_2\colon B'\rightarrow A''C$ by 
\[\CN_2\colon \tau\mapsto \CN(\Psi_1\otimes\tau).\]
In contrast to the interpretation of $\CN_1$, this may be viewed as the channel 
which would be seen by Bob if Alice were to input the $A'$ part of the purification $\ket{\Psi_2}^{A''A'}$ of $\rho_2^{A'}$ to her input of the channel and then send the $A''$ system to Charlie via a noiseless quantum channel.  As in the proof of Theorem~\ref{theo:cq}, Charlie will first decode Alice's information, after which he will attempt to simulate the channel $\CN_2$, allowing a higher transmission rate for Bob than if Alice's information was treated as noise.  Since quantum information cannot be copied, showing that this is indeed possible will require different techniques than were utilized in the previous coding theorem.
Although ensembles of random codes will be used in this proof, we introduce the technique of \emph{coherent coding}, in which we pretend that the common randomness is purified.  
The main advantage of this approach will be that working with states in the enlarged Hilbert space allows monotonicity to be easily exploited in order to provide the estimates we require.  
Additionally, before we  derandomize at the end of the proof, it will ultimately be only Bob who is using a random code.  Alice will be able to use any deterministic code from her random ensemble, as Charlie will implement a decoding procedure which produces a global state which is close to that which would have been created had Alice coded with the coherent randomness.  To show this, we will first analyze the state which would result if both senders used their full ensembles of codes.  Then we show that if Alice uses any code from her ensemble, Charlie can create the proper global state himself, allowing him to effectively simulate $\CN_2$ and ultimately decode both states at the desired rates.

By Proposition~\ref{prop:lsd}, for large enough $n$, there exists a $(2^{nR},n,\epsilon)$ random entanglement transmission code $(p_\ell,\CE_1^\ell,\CD_1^\ell)$ for the channel $\CN_1,$ where  $R = I_c(\rho_1,\CN_1)-\delta = I_c(A''\,\rangle C)-\delta.$ 
There similarly exists a $(2^{nS},n,\epsilon)$ random entanglement transmission code $(q_m,\CE_2^m,\CD_2^m)$ for $\CN_2$, with $S = I_c(\rho_2,\CN_2)-\delta = I_c(B''\,\rangle A''C) - \delta$.  
Proposition~\ref{prop:lsd} further guarantees that these codes can be chosen so that their 
respective average code density operators 
\[\varrho_1^{A'^n} = \sum_\ell p_\ell \CE_1^\ell(\pi^{\tilde{A}}) \,\text{ and }\, \varrho_2^{B'^n} = \sum_m q_m \CE_2^m(\pi^{\tilde{B}})\] satisfy
\begin{eqnarray}
|\varrho_i - \rho_i^{\otimes n}|_1\leq \epsilon \label{vrhorho}
\end{eqnarray}
and also that we may choose isometric extensions $\CU^\ell_{\CD_1}\colon C^n\rightarrow \widehat{A}F$ implementing the $\CD_1^\ell$ from Alice's random code
which satisfy
\begin{eqnarray}
F\left(\ket{\Phi_1}^{A\widehat{A}}\ket{\lambda}^{FE'^n},\CU^\ell_{\CD_1}\circ\CU_{\CN_1}^{\otimes n}\circ\CE_1^\ell\ket{\Phi_1}^{A\tilde{A}}\right) \geq 1-\epsilon \label{UD1lcond}
\end{eqnarray}
for every random code index $\ell$ and the same fixed state $\ket{\lambda}^{FE'^n}$. 

Let the code common randomness between Alice and Charlie be held between the systems $L_A$ and $L_C$, represented by the state  
\[\gamma_1^{L_AL_C} = \sum_\ell p_\ell \proj{\ell}^{L_A}\otimes \proj{\ell}^{L_C},\]
defining a similar state $\gamma_2^{M_BM_C}$ for the Bob-Charlie common randomness.  For convenience, let us further pretend that $\gamma_1$ is part of a pure state 
\[\ket{\Gamma_1}^{L_EL_AL_B} = \sum_\ell\sqrt{p_\ell} 
\ket{\ell}^{L_E}\ket{\ell}^{L_A}\ket{\ell}^{L_C}.\]
Similarly, let $\gamma_2$ by purified by $\ket{\Gamma_2}^{M_EM_BM_C}$.
Let us define the controlled encoding isometries 
$\CE_1\colon L_A\tilde{A}\rightarrow~L_AA'^n$ and $\CE_2\colon M_B\tilde{B}\rightarrow M_BB'^n$ as 
\[\CE_1 = \sum_\ell \proj{\ell}^{L_A}\otimes \CE_1^\ell 
\,\text{  and  }\,
\CE_2 = \sum_m \proj{m}^{M_B}\otimes \CE_2^m.\]
The states which would arise if Alice and Bob each encoded \emph{coherently} are
\begin{eqnarray*}
\ket{\Upsilon_1}^{LAA'^n} &\equiv& \CE_1\ket{\Gamma_1}\ket{\Phi_1} = 
\sum_\ell \sqrt{p_\ell}\ket{\ell}^L\otimes\CE_1^\ell\ket{\Phi_1} \\ 
\ket{\Upsilon_2}^{MBB'^n} &\equiv& \CE_2\ket{\Gamma_2}\ket{\Phi_2} = 
\sum_m \sqrt{q_m} \ket{m}^M\otimes\CE_2^m\ket{\Phi_2}.
\end{eqnarray*}
Note that we have abbreviated $L = L_EL_AL_C$ and $M = M_EM_BM_C$.
As each $\ket{\Upsilon_i}$ is a purification of $\varrho_i$, together with (\ref{vrhorho}), Uhlmann's theorem tells us that there exist unitaries $V_1\colon LA\rightarrow A''^n$ and $V_2\colon MB\rightarrow B''^n$ such that 
\begin{eqnarray}
F\left(V_i\ket{\Upsilon_i},\ket{\Psi_i}^{\otimes n}\right) \geq 1-\epsilon. \label{FUV}
\end{eqnarray}
Further define a corresponding controlled isometric decoder $\CU_{\CD_1}\colon L_CC^n\rightarrow L_C \widehat{A}F$ for Alice's code as
\[\CU_{\CD_1} = \sum_\ell \proj{\ell}^{L_C}\otimes \CU^{\ell}_{\CD_1}.\]
Let us now imagine that each of Alice and Bob encodes using the coherent common randomness, resulting in a global pure state 
$\CU_\CN^{\otimes n}\ket{\Upsilon_1}\ket{\Upsilon_2}$ on $LAMBC^nE^n$.  
If Charlie then applies the full controlled decoder from Alice's code, the resulting global pure state would be  
\[\ket{\Theta}^{LA\widehat{A}MBFE^n} =  
\CU_{\CD_1}\circ\CU_{\CN}^{\otimes n}
\ket{\Upsilon_1}\ket{\Upsilon_2}.\]
For each $\ell$, let us define an isometry 
$\CO^\ell\colon B'^n\rightarrow A\widehat{A}FE^n$ as
\[\CO^\ell = \CU^\ell_{\CD_1}\circ\CU_{\CN}^{\otimes n}\circ\CE_1^\ell(\Phi_1\otimes\cdot\,)\]
which we use to define the pure states 
\[\ket{\theta_\ell}^{A\widehat{A}MFBE^n} =
\CO^\ell\ket{\Upsilon_2}.\] 
These definitions allow us to express
\[\ket{\Theta} = \sum_\ell \sqrt{p_\ell} \ket{\ell}^L  \ket{\theta_\ell}.\]
Further writing $\ket{\lambda'}^{FMBE^n} \equiv V_2^{-1}\ket{\lambda}^{FB''^nE^n},$
the following bound applies
\begin{eqnarray}
F\left(\ket{\Phi_1}^{A\widehat{A}}\ket{\lambda'}^{FMBE^n},
  \ket{\theta_\ell}\right) 
&=& F\left(\ket{\Phi_1}\ket{\lambda'}^{FMBE^n},
  \CO^\ell\ket{\Upsilon_2}\right) \nonumber \\
&=& F\left(\ket{\Phi_1}\ket{\lambda}^{FB''^nE^n},
  V_2 \circ\CO^\ell\ket{\Upsilon_2} \right) \nonumber \\
&\geq& 1 - 2\sqrt{1-F\left(\ket{\Phi_1}\ket{\lambda}^{FB''^nE^n},
  \CO^\ell\ket{\Psi_2}^{\otimes n}  \right)}  \nonumber \\ 
  & & \,\,\,
  - 2\sqrt{1- F\left(V_2\ket{\Upsilon_2},\ket{\Psi_2}^{\otimes n} 
  \right)} \nonumber \\
&\geq& 1-2\sqrt{1-F\left(\ket{\Phi_1}\ket{\lambda}^{FE'^n},
  \CU^\ell_{\CD_1}\circ\CU_{\CN_1}^{\otimes n}\circ\CE_1^\ell\ket{\Phi_1}\right)} - 2\sqrt{\epsilon} \nonumber \\
&\geq& 1-4\sqrt{\epsilon}. \nonumber
\end{eqnarray}
Above, the first inequality is by the triangle inequality and monotonicity with respect to $\CO^\ell,$ while for the second inequality, we have just rewritten the first term and used (\ref{FUV}) for the second. 
The last bound is from (\ref{UD1lcond}).
Observe that we are still free to specify the global phases of the outputs of the $\CU_{\CD_1}^\ell$ so that the above bound further implies $\bra{\theta_\ell}\ket{\Phi_1}\ket{\lambda'} \geq (1-4\sqrt{\epsilon})^{1/2}$ for each $\ell$.   
Consequently, 
\begin{eqnarray}
F(\ket{\Theta},\ket{\Gamma_1}\ket{\Phi_1}\ket{\lambda'}) 
&=& \left|\sum_{\ell\ell'} \sqrt{p_\ell p_{\ell'}}\bra{\ell}\ket{\ell'}
\bra{\theta_\ell}\ket{\Phi_1}\ket{\lambda'}\right|^2 \nonumber \\
&=& \left|\sum_{\ell}p_\ell \bra{\theta_\ell}\ket{\Phi_1}\ket{\lambda'}\right|^2 \nonumber \\
&\geq& 1-4\sqrt{\epsilon}. \nonumber 
\end{eqnarray}
Essentially, the subsystems $L$, $A\widehat{A}$ and $MBFE^n$ of $\ket{\Theta}$ are mutually decoupled.

As mentioned earlier, it will be sufficient for Alice to use \emph{any} deterministic code from the random ensemble to encode.  Without loss of generality, we assume that Alice chooses to use the first code $(\ell = 1)$ in her ensemble.  Bob, on the other hand, will need to use randomness to ensure that Alice's effective channel is close to a product channel.  
The state on $AMBC^n$ which results from these encodings is $\CN^{\otimes n}(\CE^1_1(\Phi_1)\otimes\Upsilon_2)$.

We will now describe a procedure by which Charlie first decodes Alice's information, then produces a state which is close to $\Theta$, making it look like Alice had in fact utilized the coherent coding procedure.  This will allow Charlie to apply local unitaries to effectively simulate the channel $\CN_2$ for which Bob's random code was designed, enabling him to decode Bob's information as well.   These steps will constitute Charlie's decoding 
$\CD:M_CC^n\rightarrow M_C\widehat{A}\widehat{B}$, which depends on the Bob-Charlie common randomness.  The existence of a deterministic decoder will then be inferred. 

Charlie first applies the isometric decoder $\CU_{\CD_1}^1$, placing all systems into the state $\ket{\theta_1}$.   
He then removes his local system $\widehat{A}$ (it is important that he keep $\widehat{A}$ in a safe place, as it represents the decoder output for Alice's quantum information) and replaces it with the corresponding parts of the locally prepared pure state
$\ket{\Phi_1}^{A^\circ\widehat{A}^\circ}.$  Charlie also locally prepares the state $\ket{\Gamma_1}^L$.
The resulting state 
\[\Theta' = \Gamma_1^{L}\otimes\Phi_1^{A^\circ\widehat{A}^\circ}
\otimes \tr_{A\widehat{A}}\theta_1,\]
satisfies 
\begin{eqnarray}
F(\Theta',\Theta) 
&\geq& 1 - 
\left|\tr_{A\widehat{A}}\theta_1 - \lambda'\right|_1
         - \left|\lambda' - \tr_{LA\widehat{A}}\Theta\right|_1
- 3\left(1-F\big(\ket{\Gamma}\ket{\Phi_1},\tr_{MBFE^n}\Theta\big)\right) 
\nonumber \\
&\geq& 1 - 2\sqrt{4\sqrt{\epsilon}} - 2\sqrt{4\sqrt{\epsilon}}
- 12\sqrt{\epsilon} \nonumber \\
&\geq& 1 - 9\epsilon^{1/4} 
\label{FThTh}
\end{eqnarray}
whenever $\epsilon \leq 12^{-4}$.
The first line combines Lemma 2 and the triangle inequality.  The first two estimates in the second line are from 
applying (\ref{trfide}) and monotonicity with respect to $\tr_{A\h{A}}$ and $\tr_{LA\h{A}}$ to the previous two estimates.
The last estimate in that line is 
from monotonicity with respect to the map $\tr_{MBFE^n}$ applied to the previous estimate. 
Next, Charlie will 
apply $V_1\circ\CU_{D_1}^{-1}$  to $\Theta'$
\footnote{This operation only acts on Charlie's local systems, i.e.   $V_1\circ\CU_{\CD_1}^{-1}\colon L A^\circ\widehat{A}^\circ F \rightarrow A''^nC^n$.}
in order to simulate the channel $\CN_2$.  To see that this will work, 
define $\CM\colon LA\widehat{A}FE^n\rightarrow A''^nC^n$ as
$\CM\equiv\tr_{E^n}V_1\circ\CU_{\CD_1}^{-1}$
and observe that 
by monotonicity with respect to
$\CN^{\otimes n}(\,\cdot\otimes\Upsilon_2)$ and (\ref{FUV}),
the states on $MBA''^nC^n$ satisfy
\begin{eqnarray}
F\left(\CM(\Theta),\CN_2^{\otimes n}(\Upsilon_2)\right) 
&=& F\left(V_1\circ\CN^{\otimes n}(\Upsilon_1\otimes\Upsilon_2)
,\CN^{\otimes n}(\Psi_1^{\otimes n}\otimes\Upsilon_2)\right) \nonumber \\
&\geq& F\left(V_1\ket{\Upsilon_1},\ket{\Psi_1}^{\otimes n}\right) \nonumber \\
&\geq& 1-\epsilon. \nonumber
\end{eqnarray}
We may now use the triangle inequality and monotonicity with respect to 
 $\CM$ to combine our last two estimates, yielding
\begin{eqnarray}
F\left(\CM(\Theta'),\CN_2^{\otimes n}(\Upsilon_2)\right)
&\geq& 1-2\sqrt{1 - F\left(\CM(\Theta'),\CM(\Theta)\right)} 
- 2\sqrt{1-F\left(\CM(\Theta),\CN_2^{\otimes n}(\Upsilon_2)\right)} \nonumber \\   
&\geq& 1 - 2\sqrt{9\epsilon^{1/4}} - 2\sqrt{\epsilon}\nonumber \\
&\geq& 1-7\epsilon^{1/8} \label{FMN2}
\end{eqnarray}
whenever $\ep\leq2^{-8/3}$.
We have thus far shown that Charlie's decoding procedure succeeds in simulating the channel $\CN_2^{\otimes n}$, while simultaneously recovering Alice's quantum information.  
Charlie now uses the controlled decoder 
$\CD_2\colon M_CA''^nC^n\rightarrow M_C\widehat{B}$ defined as
\[\CD_2 = \sum_m \proj{m}^{M_C}\otimes \CD_2^m\]
to decode Bob's quantum information.
This entire procedure has defined our decoder $\CD:M_CC^n\rightarrow M_C\widehat{A}\widehat{B}$ which gives rise to a global state $\Omega^{A\widehat{A}B\widehat{B}}$ representing the final output state of the protocol, averaged over Bob's common randomness.
This state satisfies
\begin{eqnarray}
F(\ket{\Phi_1},\tr_{B\widehat{B}} \Omega) &\geq& F(\Theta,\Theta') 
\nonumber \\
&\geq& 1-9\epsilon^{1/4}, \nonumber
\end{eqnarray}
because of monotonicity with respect to $\tr_{LMBFE^n}$ applied to the bound (\ref{FThTh}).  By using the triangle inequality, the fact (\ref{prop1fid}) is satisfied for each $m$, and monotonicity of the estimate (\ref{FMN2}) with respect to $\tr_M\CD_2$, the global state can further be seen to obey
\begin{eqnarray}
F\big(\ket{\Phi_2},\tr_{A\widehat{A}} \Omega\big) &=& F\big(\ket{\Phi_2},\tr_M\CD_2\circ\CM(\Theta')\big) \nonumber \\
&\geq& 1 - 
2\sqrt{1-F\big(\ket{\Phi_2},\tr_M\CD_2\circ\CN_2^{\otimes n}(\Upsilon_2))\big)} \nonumber \\
& & \,\,\, - 2\sqrt{1- 
F\big(\tr_M\CD_2\circ\CN_2^{\otimes n}(\Upsilon_2),
\tr_M\CD_2\circ\CM(\Theta')\big)} \nonumber \\
&\geq& 1 - 2\sqrt{\epsilon} - 2\sqrt{7\epsilon^{1/8}} \nonumber \\
&\geq& 1 - 7 \epsilon^{1/16} \nonumber
\end{eqnarray}
as long as $\ep\leq 2^{-16/7}$.
Along with (\ref{trfide}), a final application of Lemma 2 combines the above two bounds to give
\begin{eqnarray*}
F(\ket{\Phi_1}\ket{\Phi_2},\Omega) &\geq&
1 - \left|\Phi_1 - \tr_{B\widehat{B}}\Omega\right|_1
- 3\Big(1-F\big(\ket{\Phi_2},\tr_{A\widehat{A}}\Omega\big)\Big) \\
&\geq& 1 - 2\sqrt{9\epsilon^{1/4}} - 21 \epsilon^{1/16} \\
&\geq& 1 - 22 \epsilon^{1/16},
\end{eqnarray*}
provided that $\epsilon \leq 6^{-16}$.
Since this estimate represents an average over Bob's common randomness, there must exist a particular value $m^*$ of the common randomness so that the corresponding deterministic code is at least as good as the random one, thus concluding the coding theorem.

\qed 

\section{Strong subspace transmission and scenario equivalences}
\label{section:sstequiv}
\subsection{Scenario III - Strong subspace transmission}
\label{section:sstequiv:sst}
The criteria of scenarios I and II, both in the cq and qq cases, are directly analogous to the requirement in classical information theory that the average probability of error, averaged over all codewords, be small.  Here, we introduce a situation analogous to the stronger classical condition that the \emph{maximal} probability of error be small, or that the probability of error for \emph{each} pair of codewords be small.  There are examples of classical multiple access channels for which, when
each encoder is a deterministic function from the set of the messages to the set of input symbols, the maximal error capacity region is \emph{strictly} smaller than the average error region \cite{dueck}. However, it is known that if stochastic encoders are allowed (see Problem 3.2.4 in \cite{ck}), the maximal and average error capacity regions are equal.  

It is well-known that randomization is not necessary for such an equivalence to hold for single-user channels, as Markov's inequality implies that a fraction of the codewords with the worst probability of error can be purged, while incurring a negligible loss of rate.  The obstacle to utilizing such an approach for classical multiple access channels, and hence for quantum ones as well, is that there is no guarantee that a large enough subset of bad pairs of codewords decomposes as the product of subsets of each sender's codewords.

As mentioned earlier in Section~\ref{section:background}, a particularly attractive feature of the following two scenarios is their \emph{composability}; when combined with other protocols satisfying analogous criteria, the joint protocol will satisfy similar properties.

\paragraph{III - classical-quantum scenario}
Strong subspace transmission can be considered a more ambitious version of entanglement transmission, whereby rather than requiring Bob to transmit half of a maximally entangled state $\ket{\Phi}^{B\tilde{B}},$
it is instead required that he faithfully transmit the $\tilde{B}$ part, presented to him, of \emph{any} bipartite pure state $\ket{\Psi}^{R\tilde{B}},$ where $|R|$ can be any finite number.  The reader should note that this constitutes a generalization of the usual subspace transmission \cite{bkn}, as whenever $\ket{\Psi}^{R\tilde{B}} = \ket{\psi}^R\ket{\varphi}^{\tilde{B}}$, this amounts to requiring that $\ket{\varphi}$ be transmitted faithfully.  We further demand that the maximal error probability for the classical messages be small.

As with entanglement transmission, Alice will send classical information at rate $r$ by preparing one of $2^{nr}$ pure states $\{\ket{\phi_m}^{A'^n}\}_{m\in 2^{nr}}$.  As previously discussed, our more restrictive information transmission constraints can only be met by allowing Alice to employ a stochastic encoding.  We assume that Alice begins by generating some randomness, modeled by the random variable $X$.  To send message $M=m$, she prepares the state $\ket{f(m)}\equiv\ket{\phi_{f(m)}}$, where $f(m)\equiv f_{X}(m)$ is a random encoding function, depending on the randomness in $X$.

Bob will apply an encoding $\CE\colon \tilde{B}\rightarrow B'^n,$
and Charlie will employ a decoding instrument $\boldsymbol{\CD}\colon C^n\rightarrow \widehat{M}\widehat{B}$.
These will be constructed by adding an additional layer of processing on top of the entanglement transmission codes
which were proved to exist in the previous section.
The success probability for the protocol, conditioned on $m$ being sent and $\ket{\Psi}^{R\tilde{B}}$ being presented, can be expressed as
\begin{eqnarray*}
P_s^{\text{III}}(m,\ket{\Psi}) &=& F\left(\ket{f(m)}^{\widehat{M}}\ket{\Psi}^{R\widehat{B}},
\boldsymbol{\CD}\circ\CN^{\otimes n}
\big(\phi_{f(m)}^{A'^n}\otimes\CE(\Psi^{R\tilde{B}})\big)\right).
\end{eqnarray*}
We will say that
$(f,X,\{\ket{\phi_m}\}_{m\in 2^{nr}},\CE,\boldsymbol{\CD})$ is a $(2^{nr},2^{nS},n,\epsilon)$ \emph{cq strong subspace transmission code} for the channel $\CN$ if, for every $m\in 2^{nr}$ and every $\ket{\Psi}^{R\tilde{B}}$,
\begin{eqnarray}
\E_{X}P_s^{\text{III}}(m,\ket{\Psi}) \geq 1-\epsilon.
\end{eqnarray}
The rate pair $(r,S)$ is an \emph{achievable cq rate pair for strong subspace transmission} if there is a sequence of $(2^{nr},2^{nS},n,\epsilon_n)$ cq random strong subspace transmission codes with $\epsilon_n\rightarrow 0$, and the capacity region $\CC\CQ_{\text{III}}(\CN)$ is closure of the collection of all such achievable rates.

\paragraph{III - quantum-quantum scenario}
This scenario is the obvious combination of the relevant concepts from the previous scenario and the qq entanglement transmission scenario.  Alice and Bob are respectively presented with the $\tilde{A}$ and $\tilde{B}$ parts of some pure bipartite states $\ket{\Psi_1}^{Q\tilde{A}}$ and $\ket{\Psi_2}^{R\tilde{B}}$.  As before, we place no restriction on $|Q|$ and $|R|$, other than that they are finite.
They employ their respective encodings
$\CE_1$ and $\CE_2$, while Charlie decodes with $\CD$.  As in the above cq case, the structure of these maps will be more complicated than in the previous two scenarios.
$(\CE_1,\CE_2,\CD)$ is then a $(2^{nR},2^{nS},n,\epsilon)$ \emph{qq strong subspace transmission code} if
\begin{eqnarray}
F\left(\ket{\Psi_1}^{Q\widehat{A}}\ket{\Psi_2}^{R\widehat{B}},
\CD\circ\CN^{\otimes n}\circ(\CE_1\otimes\CE_2)(\Psi_1^{Q\tilde{A}}\otimes\Psi_2^{R
\tilde{B}})\right)
\geq 1-\epsilon,
\end{eqnarray}
for every pair of pure bipartite states $\ket{\Psi_1}^{Q\tilde{A}}$ and $\ket{\Psi_2}^{R\tilde{B}}$.  Achievable rates and the capacity region $\CQ_{\text{III}}(\CN)$ are defined as in the cq case.

\subsection{Entanglement transmission implies entanglement generation}\label{section:sstequiv:eteg}
\paragraph{Proof that $\CC\CQ_{\text{II}} \subseteq \CC\CQ_{\text{I}}$: }
Suppose there exists a $(2^{nr},2^{nS},n,\epsilon)$ cq entanglement transmission code, consisting of classical message states
$\{\ket{\phi_m}^{A'^n}\}_{m\in 2^{nr}},$ a quantum encoding map $\CE\colon \tilde{B}\rightarrow \widehat{B}$, and a decoding instrument
$\boldsymbol{\CD}\colon C^n\rightarrow \widehat{M}\widehat{B}.$
Write any pure state decomposition of the encoded state
\[(1^B\otimes\CE)(\Phi) = \sum_i p_i \proj{\Upsilon_i}.\]
Then, the success condition (\ref{cqIIs}) for a cq entanglement transmission code can be rewritten as
\begin{eqnarray}
1-\epsilon &\leq& 2^{-nr}\sum_{m=1}^{2^{nr}} P_s^{\text{II}}(m) \\
&=& 2^{-nr}\sum_{m=1}^{2^{nr}}
 F\Big(\ket{\Phi}^{B\widehat{B}},
\CD_m\circ\CN^{\otimes n}\big(\phi_m^{A'^n}\otimes
\big(\sum_i p_i \Upsilon_i\big)\big)\Big) \\
&=& \sum_i p_i \left(2^{-nr}\sum_{m=1}^{2^{nr}}
F\left(\ket{\Phi}^{B\widehat{B}},
\CD_m\circ\CN^{\otimes n}(\phi_m^{A'^n}\otimes\Upsilon_i)\right)\right) \\
&=& \sum_i p_i \left(2^{-nr}\sum_{m=1}^{2^{nr}} P_s^{\text{I}}\big(m,\ket{\Upsilon_i}\big)\right),
\end{eqnarray}
so that there is a particular value $i^*$ of $i$ for which
\[2^{-nr}\sum_{m=1}^{2^{nr}} P_s^{\text{I}}\big(m,\ket{\Upsilon_{i^*}}\big)\geq 1-\epsilon.\]
Hence, $\left(\{\ket{\phi_m}\}_{m\in 2^{nr}},\ket{\Upsilon_{i^*}},\boldsymbol{\CD}\right)$ comprises an
$(2^{nr},2^{nS},n,\epsilon)$ cq entanglement generation code.

\qed

\paragraph{Proof that $\CQ_{\text{II}} \subseteq \CQ_{\text{I}}: $}
Suppose there exists a $(2^{nR},2^{nS},n,\epsilon)$ entanglement transmission code  $\left(\CE_1,\CE_2, \CD\right)$ which transmits the maximally entangled states $\ket{\Phi_1}, \ket{\Phi_2}$.  As in the cq case, the encoded states can be decomposed as
\[(1^A\otimes \CE_1)(\Phi_1) = \sum_i p_i\Upsilon_{1i}\]
and
\[(1^B\otimes \CE_2)(\Phi_2) = \sum_j q_j\Upsilon_{2i}.\]
The reliability condition (\ref{qqIIs}) can then be rewritten as
\[\sum_{ij} p_iq_jF(\ket{\Phi_1}\ket{\Phi_2},
\CD\otimes\CN^{\otimes n}(\Upsilon_{1i}\otimes\Upsilon_{2j}))
\geq 1-\epsilon,\]
which implies the existence of a particular pair $(i^*,j^*)$ of values of $(i,j)$ such that
\[F(\ket{\Phi_1}\ket{\Phi_2},
\CD\otimes\CN^{\otimes n}(\Upsilon_{1i^*}\otimes\Upsilon_{2j^*}))
\geq 1-\epsilon.\]
Hence, $\left(\ket{\Upsilon_{1i^*}},\ket{\Upsilon_{2j^*}},\CD\right)$ comprises a $(2^{nR},2^{nS},n,\epsilon)$ qq entanglement generation code.

\qed

\subsection{Entanglement transmission implies strong subspace transmission} \label{section:sstequiv:etsst}
\paragraph{Proof that $\CC\CQ_{\text{II}} \subseteq \CC\CQ_{\text{III}}$:}
Suppose there exists a $(2^{nr},2^{nS},n,\epsilon^2/2)$ entanglement transmission codes with classical message states $\{\ket{\phi_m}^{A'^n}\}_{m\in 2^{nr}},$ quantum encoding $\CE\colon \tilde{B}\rightarrow \widehat{B},$ and decoding instrument $\boldsymbol{\CD}\colon C^n\rightarrow \widehat{M}\widehat{B}$ with trace-reducing components $\{\CD_m:C^n\rightarrow \widehat{B}\}$.

We will initially prove the equivalence by constructing a code which requires  two independent sources of shared common randomness $X$ and $Y$.  $X$ is assumed to be available to Alice and to Charlie, while $Y$ is available to Bob and to Charlie.  Then, we will argue that it is possible to eliminate the dependence on the shared randomness, by using the channel to send a neglibly small ``random seed", which can be recycled to construct a code which asymptotically achieves the same performance as the randomized one.

We begin by demonstrating how shared common randomness between Alice and Charlie allows Alice to send any message with low probability of error.  Setting $\mu=2^{nr}$, let the random variable $X$ be uniformly distributed on the set $\{1,\dotsc,\mu\}$.  To send message $M=m$, Alice computes $m' = m + X$ modulo $\mu$.  She then prepares the state $\ket{\phi_{m'}}$ for transmission through the channel.  Bob encodes the $\tilde{B}$ part of $\ket{\Phi}^{B\tilde{B}}$ with $\CE$, and each sends appropriately through the channel.  Charlie decodes as usual with the instrument $\boldsymbol{\CD}$.  Denoting the classical output as $\widehat{M}'$, his declaration of Alice's message is then
$\widehat{M} = \widehat{M}' - X$ modulo $\mu$.
Defining the trace-reducing maps
$\CM_m\colon \tilde{B}\rightarrow \widehat{B}$ by
\[\CM_m\colon \tau \mapsto
\CD_m\circ\CN^{\otimes n}(\phi_m\otimes\CE(\tau)),\]
and the trace-reducing average map as
\[\CM\colon\tau\rightarrow \frac{1}{\mu}\sum_{m=1}^\mu \CM_m(\tau),\]
we can rewrite the success criterion (\ref{cqIIs}) for entanglement transmission as
\begin{eqnarray*}
F(\ket{\Phi},\CM(\Phi)) \geq 1-\epsilon^2/2,
\end{eqnarray*}
which, together with (\ref{trfide}), implies that for the identity map $\text{id}:\tilde{B}\rightarrow \widehat{B}$,
\begin{eqnarray}
\left|(\CM - \text{id})(\Phi)\right|_1 \leq \epsilon. \label{cqsim}
\end{eqnarray}
The above randomization of the classical part of the protocol can be mathematically expressed by replacing the $\CM_m$ with $\CM_{m+X}$.  As tracing over the common randomness $X$ is equivalent to computing the expectation with respect to $X$, we see that $\E_X\CM_{m+X} = \CM$, or rather
\[\E_X F(\ket{\Phi},\CM_{m+X}(\Phi)) = F(\ket{\Phi},\CM(\Phi)).\]
It is thus clear that the maximal error criterion for the randomized protocol is equal to the average criterion for the original one.

We continue by randomizing the quantum part of the classically randomized protocol.
Setting $d = 2^{nS} = |\tilde{B}|,$
let $\{U_y\}_{y \in d^2}$ be the collection of Weyl unitaries, or
generalized Pauli operators, on the $d$-dimensional input space.
Observe that for any $\rho$, acting with a uniformly random choice of Weyl unitary has a completely randomizing effect, in the sense that
\[\frac{1}{d^2}\sum_{y=1}^{d^2} U_y\rho U_y^{-1} = \pi_d.\]
Let the random variable $Y$ be uniformly distributed on $\{1,\dotsc,d^2\}$. It will be convenient to define the common randomness state
\[\Upsilon^{Y_BY_C} = \frac{1}{d^2}\sum_{y=1}^{d^2} \proj{y}^{Y_B}\otimes\proj{y}^{Y_C},\]
where the system $Y_B$ is in the possession of Bob, while $Y_C$ is possessed by
Charlie.  Define now the controlled unitaries
$\CU_B\colon Y_B \tilde{B}\rightarrow Y_B\tilde{B}$ and
$\CU_C\colon Y_C \widehat{B}\rightarrow Y_C\widehat{B}$ by
\[\CU_B =
  \sum_{y=1}^{d^2} \proj{y}^{Y_B}\otimes U_y\]
and
\[\CU_C =
  \sum_{y=1}^{d^2} \proj{y}^{Y_C}\otimes U_y^{-1}.\]
Suppose Bob is given the $\tilde{B}$ part of an arbitrary pure state $\ket{\Psi}^{R\tilde{B}}$, where $|R| < \infty,$ and Alice sends the classical message $M=m$.
For encoding, Bob will apply $\CE\circ\CU_B$ to the combined system $\Upsilon\otimes\Psi$.  Charlie decodes with $\CU_C\circ\CD$.  If $\CM$ were equal to the perfect quantum channel $\text{id}\colon\tilde{B}\rightarrow\widehat{B}$, this procedure would result in the state
\[\frac{1}{d^2}\sum_{y=1}^{d^2} \proj{y}^{Y_B}\otimes\proj{y}^{Y_C}
\otimes \Psi.\]
Note that the common randomness is still available for reuse.
Abbreviating $\proj{y}^Y = \proj{y}^{Y_B}\otimes\proj{y}^{Y_C}$,
and  $\ket{\Psi_y}^{R\tilde{B}} = (1^R\otimes U_y)\ket{\Psi}$, we write
\begin{eqnarray}
\sigma^{YR\tilde{B}} &=& \CU_B(\Upsilon\otimes\Psi) \\  &=&\frac{1}{d^2}\sum_{y=1}^{d^2}\proj{y}^Y\otimes\Psi_y.
\end{eqnarray}
Observe that  $\sigma$ is an extension of the maximally mixed state $\pi^{\tilde B}$, and can be seen to arise by storing in $Y$ the result of a von Neumann measurement along the basis $\{\ket{y}^{R'}\}_{y\in d^2}$
on the $R'$ part of the pure state
\[\ket{\Gamma}^{R'R\tilde{B}} =
  \frac{1}{d}\sum_{y=1}^{d^2}\ket{y}^{R'}\ket{\Psi_y}^{R\tilde{B}}.\]
Since $\tr_{R'R} \Gamma = \tr_{YR}\sigma = \pi^{\tilde{B}}$, $\ket{\Gamma}$ is maximally entangled between $R'R$ and $\tilde{B}$.  So, there exists an isometry $V\colon B\rightarrow R'R$ such that
$(V\otimes 1^{\tilde{B}})\ket{\Phi}^{B\tilde{B}} = \ket{\Gamma}.$  This implies that there is a quantum operation $\CO\colon B\rightarrow YR$  such that $(\CO\otimes 1^{\tilde{B}})(\Phi) = \sigma$.
Define the trace-reducing map $\CT\colon\tilde{B}\rightarrow \widehat{B},$ which represents the coded channel with common randomness accounted for, by
\[\CT\colon\tau\mapsto \tr_{Y} \CU_C\circ\CM\circ\CU_B(\Upsilon\otimes\tau).\]
Recalling our denotation of the noiseless quantum channel $\text{id}\colon\tilde{B}\rightarrow \widehat{B}$, 
as well as our convention that id acts as the identity on any system which is not $\tilde{B}$, we now bound
\begin{eqnarray*}
1-F(\ket{\Psi},\CT(\Psi))
&\leq& \big|(\CT - \text{id})
  (\Psi)\big|_1 \\
&\leq& \big|(\CU_C\circ\CM\circ\CU_B - \text{id})
  (\Upsilon\otimes\Psi)\big|_1 \\
&=&\big|(\CM - \text{id})\circ\CU_B(\Upsilon\otimes\Psi)\big|_1 \\
&=&\big|(\CM - \text{id})(\sigma)\big|_1 \\
&\leq& \big|(\CM - \text{id})(\Phi)\big|_1 \\
&\leq& \epsilon,
\end{eqnarray*}
where the first line is by (\ref{trfid}) and the second by monotonicity with respect to $\tr_Y$.  The third follows from unitary invariance of the trace.  The second to last inequality is a consequence of monotonicity with respect to $\CO$, while the last is by (\ref{cqsim}).  Note that by monotonicity, this implies that any density matrix $\Omega^{R\tilde{B}}$ satisfies
\begin{eqnarray}
|\CT(\Omega) - \Omega|_1\leq \epsilon. \label{TOmega}
\end{eqnarray}

We have thus shown that if Alice and Charlie have access to a common randomness source of rate $r$, while Bob and Charlie can access one of rate $2S$, the conditions for strong subspace transmission can be satisfied.  Next, we will illustrate that, by modifying our protocol, it is possible to reduce the amount of shared randomness required.  Using the previous blocklength-$n$ construction, we will concatenate $N$ such codes, where each utilizes the \emph{same} shared randomness, to construct a new code with blocklength $nN$.
For an arbitrary $\ket{\Psi^{(N)}}^{R\tilde{B}^N}$, further define
the commuting operations $\{\CT_i\}_{i\in N},$ where $\CT_i\colon\tilde{B}_i\rightarrow \widehat{B}_i$ is $\CT$ acting on the $i$'th tensor factor of $\Psi^{(N)}.$  Setting $\xi_0\equiv\Psi^{(N)}$,  we then recursively define the density operators $\xi_i = \CT_i(\xi_{i-1}),$ noting that
$\xi_N = \CT_N\circ\cdots\circ\CT_1(\xi_0) = \CT^{\otimes N}(\Psi^{(N)})$.
Because of (\ref{TOmega}),
$|\xi_{i+1} - \xi_{i}|_1 = |\CT_{i+1}(\xi_i) - \xi_i|_1 \leq \epsilon$,
and we can use the triangle inequality to estimate
\begin{eqnarray*}
\big|\CT^{\otimes N}(\Psi^{(N)}) - \Psi^{(N)}\big|_1
&=& \big|\xi_{N} - \xi_{0}\big|_1 \\
&\leq&
  \sum_{i=1}^N \big|\xi_i - \xi_{i-1} \big|_1 \\
&\leq& N\epsilon.
\end{eqnarray*}
By choosing $N = \frac{1}{\sqrt{\epsilon}}$, it is clear that we have reduced Alice's and Bob's shared randomness rates respectively to $\sqrt{\epsilon}r$ and $2\sqrt{\epsilon}S$, while the error on the $N$-blocked protocol is now $\sqrt{\epsilon}$.  Next, we argue that by using two more blocks of length $n$, it is possible to simulate
the shared randomness by having Alice send $nr$ random bits $X$ using the first block, while Bob locally prepares two copies of $\ket{\Phi}$, written $\ket{\Phi}^{B_1\tilde{B}_1}\ket{\Phi}^{B_2\tilde{B}_2}$,
and transmits the $\tilde{B}_1\tilde{B}_2$ parts over the channel using both blocks.
Charlie decodes each block separately, obtaining a random variable $\widehat{X}$ and the $\widehat{B}_1$ and $\widehat{B}_2$ parts of the
post-decoded states $\Omega_1^{B_1\widehat{B_1}}$ and $\Omega_2^{B_2\widehat{B_2}}.$
Bob and Charlie then measure their respective parts of $\Omega_1\otimes\Omega_2$ in some previously agreed upon orthogonal bases to obtain a simulation $\widehat{\Upsilon}$ of the perfect shared randomness state which, by monotonicity and telescoping, satisfies
\begin{eqnarray*}
|\Upsilon - \widehat{\Upsilon}|_1 &\leq& |\Phi\otimes\Phi - \Omega_1\otimes\Omega_2|_1 \\
&\leq& \epsilon^2.
\end{eqnarray*}
Further, the noisy shared randomness for the classical messages can be shown to satisfy
\begin{eqnarray*}
\big|\dist(X,X) - \dist(X,\widehat{X})\big|_1 &=& 2\Pr\{X=\widehat{X}\} \\
&\leq& \epsilon^2.
\end{eqnarray*}
By monotonicity of trace distance and the triangle inequality, using the noisy common randomness state $\widehat{\Upsilon}$ increases the estimate for each block by $2\epsilon^2$.  For identical reasons, the same increase is incurred by using the noisy common randomness $(X,\widehat{X})$.  Thus, accounting for both sources of noisy common randomness,
the estimate (\ref{TOmega}) is changed to $2\epsilon$, provided that $\epsilon\leq \frac{1}{4}$.
The noisy common randomness thus increases the bound on the error of the $N$-blocked protocol to $2\sqrt{\epsilon}$, while costing each of Alice and Bob a negligible rate overhead of $\frac{2}{N+2}$ in order to seed the protocol.

The above protocol can be considered as defining an encoding map $\CE'\colon\tilde{B}^N\rightarrow B'^{(N+2)n}$
and decoding instrument
$\boldsymbol{\CD}\colon C^{(N+2)n}\rightarrow \widehat{B}^N \widehat{M}^N$.
Thus, the protocol takes a $(2^{nr},2^{nS},n,\epsilon_n)$ cq entanglement transmission code and constructs a $(2^{n'r'},2^{n'S'},n',\epsilon'_{n'})$ strong subspace transmission code with cq rate pair $(r',S') = \left(\frac{r}{1+\epsilon'_{n'}},\frac{S}{1+\epsilon'_{n'}}\right),$
where $n' = \left(2+\frac{1}{\sqrt{\epsilon_n}}\right)n$, and $\epsilon'_{n'} = 2\sqrt{\epsilon_n}$.
Now, if the rates $(r,S)$ are achievable cq rates for entanglement
transmission, there must exist a sequence of $(2^{nr},2^{nS},n,2\epsilon_n^2)$ entanglement transmission codes with $\epsilon_n\rightarrow 0$.  Since this means that $\frac{1}{1+2\sqrt{\epsilon_n}}$ increases to unity, we have shown that for any $\delta > 0$, every rate pair $(r-\delta,S-\delta)$ is an achievable cq rate pair for strong subspace transmission.  Since the capacity regions for each scenario are defined as the closure of the achievable rates, this completes the proof.

\qed

\paragraph{Proof that $\CQ_{\text{II}} \subseteq \CQ_{\text{III}}$: }
We will employ similar techniques as were used in the previous proof
to obtain this implication.
Suppose there exists a $(2^{nR},2^{nS},n,\frac{1}{2}\epsilon^2)$ qq entanglement transmission code $(\CE_1,\CE_2,\CD)$, with
$\CE_1\colon\tilde{A}\rightarrow A'^n$, $\CE_2\colon\tilde{B}\rightarrow B'^n$, and $\CD\colon C^n\rightarrow \widehat{A}\widehat{B}.$
Setting $a = |\tilde{A}|= 2^{nR}$ and $b = |\tilde{B}|=2^{nS}$,
define the common randomness states
\[\Upsilon^{X_AX_C}_X = \frac{1}{a^2}\sum_{x=1}^{a^2}\proj{x}^{X_A}\otimes\proj{x}^{X_C}\]
and
\[\Upsilon^{Y_BY_C}_Y = \frac{1}{b^2}\sum_{x=1}^{b^2}\proj{y}^{Y_B}\otimes\proj{y}^{Y_C}\]
These states will be used as partial inputs to the controlled unitaries
\begin{eqnarray*}
\CU_A &=&
 \sum_{x=1}^{a^2}\proj{x}^{X_A}\otimes U_x,\\
\CU_C &=&
 \sum_{x=1}^{a^2}\proj{x}^{X_C}\otimes U_x^{-1},\\
\CV_B &=&
 \sum_{y=1}^{b^2}\proj{y}^{Y_B}\otimes V_x,\\
\CV_C &=&
 \sum_{y=1}^{b^2}\proj{y}^{Y_C}\otimes V_x^{-1}\\
\end{eqnarray*}
where, as before, we have utilized the Weyl unitaries
$\{U_x\}_{x\in a^2}$ and $\{V_y\}_{y\in b^2}$, which respectively completely randomize any states on $a$-dimensional and $b$-dimensional spaces.  Suppose Alice and Bob are respectively presented with the $\tilde{A}$ and $\tilde{B}$ parts of the arbitrary pure states $\ket{\Psi_1}^{Q\tilde{A}}$ and $\ket{\Psi_2}^{R\tilde{B}}.$
Writing $\CM = \CD\circ\CN^{\otimes n}\circ(\CE_1\otimes\CE_2)$, and defining the map $\CT\colon\tilde{A}\tilde{B}\rightarrow \widehat{A}\widehat{B}$ by
\[\CT\colon \tau \mapsto
(\CU_C\otimes\CV_C)\circ\CM\circ(\CU_A\otimes\CV_B)
(\tau\otimes\Upsilon_1\otimes\Upsilon_2),\]
the overall joint state of the randomized protocol is given by $\CT(\Psi_1\otimes\Psi_2)$.
Abbreviating
\[\proj{xy}^{XY} = \proj{x}^{X_A}\otimes\proj{x}^{X_C}
\otimes\proj{y}^{Y_B}\otimes\proj{y}^{Y_C}\]
and defining $\ket{\Psi_x}^{Q\tilde{A}} = (1^{Q}\otimes U_x)\ket{\Psi_1}$,
$\ket{\Psi_y}^{R\tilde{B}} = (1^{R}\otimes V_y)\ket{\Psi_2},$
we write
\[\sigma^{XYQR\tilde{A}\tilde{B}}
= \frac{1}{a^2 b^2}\sum_{xy} \proj{xy}\otimes \Psi_x \otimes \Psi_y.\]
By similar arguments as in the cq case, there exists a map
$\CO\colon AB\rightarrow XYQR$ so that
\[(\CO\otimes 1^{\tilde{A}\tilde{B}})(\Phi_1\otimes\Phi_2) = \sigma.\]
Again, for the same reasons as in the cq case, we have
\begin{eqnarray*}
|(\CT - \text{id})(\Psi_1\otimes\Psi_2)|_1
&\leq& |(\CM - \text{id})(\sigma)|_1 \\
&\leq& |(\CM - \text{id})(\Phi_1\otimes\Phi_2)|_1 \\
&\leq& \epsilon.
\end{eqnarray*}
The rest of the proof is nearly identical to that from the previous section, so we omit these details, so as not to have to repeat our previous arguments here.
\qed

\subsection{Strong subspace transmission implies entanglement transmission}\label{section:sstequiv:sstet}
\paragraph{Proof that $\CC\CQ_{\text{III}} \subseteq \CC\CQ_{\text{II}}$: }
Given a strong subspace transmission code, if Alice uses any deterministic value $x$ for her locally generated randomness $X$, the average classical error will be equal to the expected maximal classical error of the randomized code.  Since the ability to transmit any state includes the maximally entangled case, this completes the claim.

\paragraph{Proof that $\CQ_{\text{III}} \subseteq \CQ_{\text{II}}$: }
This implication is immediate.  As any states can be transmitted, this certainly includes the case of a pair of maximally entangled states.

\section{Discussion}\label{section:discussion}
There have been a number of results analyzing multiterminal coding problems in quantum Shannon theory.
For an i.i.d.\ classical-quantum source $XB$, Devetak and Winter \cite{dw} have proved a Slepian-Wolf-like coding theorem achieving the cq rate pair $(H(X|B),H(B))$ for classical data compression with quantum side information.  Such codes extract classical side information from $B^n$ to aid in compressing $X^n.$  The extraction of side information is done in such a way as to cause a negligible disturbance to $B^n$. Our Theorem~\ref{theo:cq} is somewhat of this flavor.  There, the quantum state of $C^n$ is measured to extract Alice's classical message which, in turn, is used as side information for decoding Bob's quantum information.  
Analogous results to ours were obtained by Winter in his analysis of a multiple access channel with classical inputs and a quantum output, whereby the classical decoded message of one sender can be used as side information to increase the classical capacity of another sender.

We further mention the obvious connection between our coding theorems and the subject of channel codes with side information available to the receiver.  
The more difficult problem of classical and quantum capacities when side information is available at the \emph{encoder} is analyzed by Devetak and Yard in \cite{dy}, constituting quantum generalizations of results obtained by Gelfand and Pinsker \cite{gp} for classical channels with side information.

In an earlier draft of this paper, we characterized $\CQ(\CN)$ as the closure of a regularized union of rectangles 
\begin{eqnarray*}
0 \!\!\!&\leq \, R  \,\leq& \!\!\!\frac{1}{k}I_c(A\,\rangle C^k)  \\
0 \!\!\!&\leq \, S  \,\leq&  \!\!\!\frac{1}{k}I_c(B\,\rangle C^k). 
\end{eqnarray*}
This solution had been conjectured on the basis of a duality between classical Slepian-Wolf distributed source coding and classical 
multiple-access channels \cite{ck,coverthomas}, as well as on a purported no-go theorem for distributed data compression of so-called irreducible pure state ensembles that appeared in an early version of \cite{adhw}.  After the earlier preprint was made available, Andreas Winter announced 
\cite{wintertalk} recent progress with Jonathan Oppenheim and Michal Horodecki \cite{how} on the quantum Slepian-Wolf problem, offering a characterization identical in functional form to the classical one, while also supplying an interpretation of negative rates and apparently evading the no-go theorem.  Motivated by the earlier mentioned duality, he informed us that the qq capacity region could also be characterized in direct analogy to the classical case.  Subsequently, we found that we could modify our previous coding theorem to achieve the new region, provided that the rates are nonnegative.  
After those events unfolded, the authors of \cite{adhw} found an error in the proof of their no-go theorem, leading to a revised version consistent with the newer developments. 
Our earlier characterization of $\CQ(\CN)$, while correct, is contained in the rate region of Theorem 2 for any finite $k$, frequently strictly so.  The newer theorem, therefore, gives a more accurate approximation to the rate region for finite $k$.  In fact, for any state arising from the channel which does not saturate the strong subadditivity inequality \cite{hjpw}, the corresponding pentagon and rectangle regions are distinct.  Another beneficial feature of the new characterization is that it is possible to show that the maximum sum rate bound $R+S \leq \max I_c(AB\,\rangle C)$ is additive, where the maximization is over all states of the form (\ref{th2arise}), for any channel which is \emph{degradable} in the sense of \cite{ds}.

More recently, we discovered that the same technique used to prove the new characterization of $\CQ(\CN)$ implies a new cq coding theorem, and thus a new characterization of $\CC\CQ(\CN)$.  By techniques nearly identical to those employed in the coding theorem for Theorem~2,
it is possible to achieve the cq rate pair 
\[(r,S) = \big(I(X;BC),I_c(B\,\rangle C)\big)\]
corresponding to Bob's quantum information being used as side information for decoding Alice's classical message.
This is accomplished by having Charlie isometrically decode Bob's quantum information, then coherently decode to produce an effective channel 
$\CN_1\colon A'\rightarrow BC$ so that Alice can transmit classically at a higher rate.  The new characterization is then a regularized union of pentagons, consisting of pairs of nonnegative rates $(r,S)$ satisfying
\begin{eqnarray*}
r &\leq& I(X;BC) \\
S &\leq& I_c(B\,\rangle CX)\\
r+S &\leq& I(X;C) + I_c(B\,\rangle CX) = I(X;BC) + I_c(B\,\rangle C).
\end{eqnarray*} 
Surprisingly, it is thus possible to characterize each of $\CC\CQ(\CN)$ and $\CQ(\CN)$ in terms of pentagons, in analogy to the original classical result.  This situation makes apparent the dangers of being satisfied with regularized expressions for capacity regions.  Without being able to prove single-letterization steps in the converses, it is hard to differentiate which characterization is the ``right" one.  While it is intuitively satisfying to see analogous formulae appear in both the classical and quantum theories, the regularized nature of the quantum results blurs the similarity.  Indeed, the problems with single-letterization for single-user channels appear to be amplified when analyzing quantum networks (see e.g.\ \cite{dch}).  Perhaps this indicates that the necessity of understanding the capacities of single-user channels at a level beyond regularized optimizations is even more pressing than previously thought.  We should mention that for the erasure channel analyzed in Section~\ref{section:appendix:cqadd}, the newer description of $\CC\CQ(\CN)$ is not an issue, as the new corner point is contained in the old rectangle for any state arising from any number of parallel instances of the erasure channel.  On the other hand, we demonstrate in \cite{isit} that for the collective phase flip channel, both characterizations single-letterize, yielding a classical-quantum region which is identical in form to that obtained for $\CQ(\CN)$ in Section~\ref{section:appendix:qqadd}, replacing the quantum rate with a classical one.

Consider the full simultaneous classical-quantum region $\CS(\CN)$ for two senders, where each sends classical and quantum information at the same time.  A formal operational definition of $\CS(\CN)$ is found in \cite{yardthesis,isit}.  This region can be characterized in a way that generalizes Theorems~1 and 2 as the regularization of the region $\CS^{(1)}(\CN)$,
defined as the vectors of nonnegative rates $(r,s,R,S)$ satisfying
\begin{eqnarray*}
r &\leq& I(X;C|Y) \\
s &\leq& I(Y;C|X) \\
r + s &\leq& I(XY;C) \\
R &\leq& I_c(A\,\rangle BCXY) \\
S &\leq& I_c(B\,\rangle ACXY) \\
R+S &\leq& I_c(AB\,\rangle CXY)
\end{eqnarray*}
for some state of the form
\[\sigma^{XYABC} =
\sum_{x,y}p(x)p(y)\proj{x}^X\otimes\proj{y}^Y
\otimes\CN(\psi_x^{AA'}\otimes\phi_y^{BB'}),\]
arising from the action of $\CN$ on the $A'$ and $B'$ parts of some pure state ensembles
$\{p(x),\ket{\psi_x}^{AA'}\}$, $\{p(y),\ket{\phi_y}^{BB'}\}$.
Briefly, achievability of this region is obtained as follows.  Using techniques introduced in \cite{ds}, each sender ``shapes" their quantum information into HSW codewords.  Decoding is accomplished by first decoding all of the classical information, then using that information as side information for a quantum decoder.  
The main result of 
\cite{ds}, the regularized optimization of the cq result from \cite{wintermac} over pairs of input ensembles, and our Theorems 1 and 2 follow as corollaries of the corresponding capacity theorem.  Indeed, the six two-dimensional ``shadows" of the above region, obtained by setting pairs of rates equal to zero, reproduce those aforementioned results.
This characterization, however, only utilizes the rectangle description of $\CC\CQ(\CN)$.  It is indeed possible to write a more accurate regularized description of $\CS(\CN)$ which generalizes the pentagon characterizations of $\CC\CQ(\CN)$ and $\CQ(\CN)$, although we will not pursue that at this time.
 
\paragraph{Acknowlegements}  JY would like to thank T.\ Cover for much useful input and feedback at many stages during the writing of this manuscript, Y.\ H.\ Kim for useful discussions regarding classical multiple access channels.  JY has been supported by the Army Research Office MURI under contract DAAD-19-99-1-0215, the National Science Foundation under grant CCR-0311633, and the Stanford Networking Research Center under grant 1059371-6-WAYTE.
PH is grateful for support from the Canadian Institute for Advanced Research, the Sherman Fairchild Foundation and the US National Science Foundation under grant EIA-0086038.

\section{Appendix}\label{section:appendix}
\subsection{Proof of additivity of $\CC\CQ$ for quantum erasure multiple access channel} \label{section:appendix:cqadd}
Due to the regularized form of our Theorems 1 and 2, the possibility of actually computing the capacity regions seems generally out of reach.  Here we give some examples of channels whose capacity region does in fact admit a single-letter characterization, in the sense that no regularization is necessary.

Our first example is a multiple access erasure channel $\CN\colon A'B'\rightarrow C$, where $|A'|=2, |B'|=d$ and $|C|=d+1.$  Alice will send classical information while Bob will send quantum.  Fixing bases $\{\ket{0}^{A'},\ket{1}^{A'}\},
 \{\ket{1}^{B'},\dotsc\ket{d}^{B'}\},
\{\ket{0}^C,\dotsc,\ket{d}^C\},$ the channel
has $d+1$ operation elements
\begin{eqnarray*}
N_0 &=& \sum_{j=1}^d\ket{0}^C\bra{0}^{A'}\bra{j}^{B'} \\
N_i &=& \ket{i}^C\bra{1}^{A'}\bra{i}^{B'}, \,\,\,\,\,i=1,\dotsc d.
\end{eqnarray*}
The action of the channel can be interpreted as follows.  First, a projective measurement of Alice's input along $\{\ket{0},\ket{1}\}$ is performed.  If the result is $0$, Charlie's output is prepared in a pure state $\ket{0}$.  Otherwise, Bob's input is transferred perfectly to the remaining degrees of freedom in Charlie's output.  Bob's input is ``erased", or otherwise ejected into the environment, whenever Alice sends $\ket{0}$, and is perfectly preserved when she sends $\ket{1}$.
Indeed, the action of $\CN$ on $\tau^{A'}\otimes\rho^{B'}$ is given by
\[\CN(\tau\otimes\rho) = \tau_{00}\proj{0} + \tau_{11}\rho.\]

In the sense of (\ref{th1arise}), any state $\Omega^{XBC^k}$ which arises from $\CN^{\otimes k}$ can be specified by fixing some pure state ensemble $\{p(x),\ket{\phi_x}^{A'^k}\}$ and a pure bipartite state $\ket{\Psi}^{BB'^k}$.  We thus write
\[\Omega = \sum_x p(x)\proj{x}^X\otimes(1^B\otimes\CN^{\otimes k})(\phi_x\otimes\Psi).\]
For a binary string $y^k$, let $\ket{y^k}^{A'^k}=\ket{y_1}^{A'}\cdots\ket{y_k}^{A'}$ be the associated computational basis state.  Writing
$p(y^k|x) = |\braket{y^k}{\phi_x}|^2$ defines the random variable $Y^k$,
which is correlated with $X$, and can be interpreted as the erasure pattern associated with the state $\Omega$.
We next define another state of the form (\ref{th1arise}),
\[
\Omega'^{XY^kBC^k} = \sum_{x,y^k} p(x)p(y^k|x)\proj{x}^X\otimes\proj{y^k}^{Y^k}\otimes
\CN^{\otimes k}(\proj{y^k}\otimes\Phi),
\]
for
\[\ket{\Phi}^{BB'^k} =  \sum_{j^k}\ket{j^k}^B\ket{j_1}^{B'_1}\cdots\ket{j_k}^{B'_n},\]
where the summation is over $d$-ary strings of length $k$, $j^k = (j_1,\dotsc,j_k).$
Finally, for
\begin{eqnarray*}
q_i &=& \Pr\{Y_i = 0\}, \\
q &=& \frac{1}{k}\sum_{i=1}^kq_i, \\
\ket{\varphi}^{BC} &=&\frac{1}{\sqrt{d}}\sum_{j=1}^d\ket{j}^B\ket{j}^{C},
\end{eqnarray*}
define a third state
\[\omega^{UBC} = q\proj{0}^U\otimes\pi_d^B\otimes\proj{0}^C
+ (1-q) \proj{1}^U\otimes\varphi^{BC}.\]
The above states can easily be seen to satisfy the following chain of inequalities
\begin{eqnarray*}
I(X;C^k)_\Omega &=& I(X;C^k)_{\Omega'} \\
&=& I(X;Y^k)_{\Omega'} \\
&\leq& H(Y^k)_{\Omega'} \\
&\leq& \sum_{i=1}^k H(Y_i)_{\Omega'} \\
&=& \sum_{i=1}^k H(q_i) \\
&\leq& k H(q) \\
&=& k H(U)_\omega \\
&=& k I(U;C)_\omega.
\end{eqnarray*}
The only nontrivial step above is that we have used the concavity of the binary entropy function in the last inequality.
Furthermore, it is not hard to see that
\begin{eqnarray*}
I_c(B\,\rangle C^kX)_\Omega 
&\leq& I_c(B\,\rangle C^kXY^k)_{\Omega'} \\
&=& I_c(B\,\rangle C^kY^k)_{\Omega'}\\
&=& kI_c(B\,\rangle CU)_\omega.
\end{eqnarray*}
Thus, we have shown that for any state $\Omega^{XBC^k}$ arising from
$\CN^{\otimes k}$ in the sense of (\ref{th1arise}), there is a state $\omega^{UBC}$ arising from $\CN$ in the same sense, allowing the multi-letter information quantities to be bounded by single-letter information quantities; i.e. $\CC\CQ(\CN) = \CC\CQ^{(1)}(\CN)$.
\qed

As it is clear that $I(U;C)_\omega = H(q)$, we focus on calculating
\begin{eqnarray*}
I_c(B\,\rangle CU)_\omega
&=& q \left(H(\proj{0}^C) - H(\pi_d^B\otimes\proj{0}^C)\right)
 + (1-q) \Big(H(\pi_d^C) - H(\varphi^{BC})\Big) \\
&=& q (0-\log d) + (1-q)(\log d - 0)\\
&=& (1-2q)\log d.
\end{eqnarray*}
Note that the above quantity is a weighted average of a positive and a  negative coherent information.  It is perhaps tempting to interpret these terms as follows.  The positive term can be considered as resulting from a preservation of quantum information, while the 
negative term can be seen as signifying a complete loss of quantum information to the environment.  The overall coherent information is positive only when $q < \frac{1}{2}$, a result which is in agreement with the result of Bennett et al.\ \cite{bds}
on the quantum capacity of a binary erasure channel.
Varying $0\leq q \leq \frac{1}{2},$
the rate pairs
\begin{eqnarray*}
(r,S) &=&\big(I(U;C),I_c(B\,\rangle CU)\big)_\omega \\
&=& \big(H(q),(1-2q)\log d\big)
\end{eqnarray*}
can be seen to parameterize the outer boundary of $\CC\CQ(\CN)$,
as is pictured in figure \ref{erasure} for the case $d=2.$

As an aside, we remark that this calculation, together with the quantum channel capacity theorem from \cite{dev}, gives a direct derivation of the quantum capacity of a quantum erasure channel, without relying on the no-cloning and hashing arguments used in \cite{bds}.

\subsection{Proof of additivity of $\CQ$ for collective phase flip channel}
\label{section:appendix:qqadd}
While the description of the capacity region $\CQ$ in Theorem~2
generally requires taking a many-letter limit, we give here an
example of a quantum multiple access channel $\CN_p\colon A'B'\rightarrow C$ for which that description can be single-letterized.
The channel $\CN_p$ takes as input two qubits, one from Alice and the other from Bob.  With probability $p$, the channel causes each qubit to undergo a phase flip, by rotating the state of each by 180$^\circ$ about its z-axis on the Bloch sphere, before it is received by the receiver Charlie.  The action of $\CN_p$ on an input density operator $\rho^{A'B'}$ is described in terms of the operator sum representation as 
\[\CN_p(\rho) = (1-p)\rho + p (\sig_z\otimes \sig_z)\rho (\sig_z \otimes \sig_z),\]
where 
$\sig_z$ is the Pauli phase flip matrix.
We will demonstrate that $\CQ(\CN_p)$ is equal to the collection of all pairs of nonnegative rates $(R,S)$ which satisfy 
\begin{eqnarray*}
 R &\leq& 1 \\
 S &\leq& 1 \\
 R + S &\leq& 2-H(p).
\end{eqnarray*}
\begin{proof}
We first observe that for any input state of the form (\ref{th2arise}),  $\frac{1}{k}I_c(AB\,\rangle C^k)$ is upper bounded by the quantum capacity of $\CN_p$ when both senders may act together.  Since $\CN_p$ is a \emph{generalized dephasing channel} \cite{ds}, its quantum capacity can be calculated by a single-letter optimization of the coherent information over input density operators which are diagonal in the dephasing basis which, incidentally, is just the computational basis.
A short calculation reveals that it suffices to check inputs of the form 
$\rho_\al = \frac{1}{2}\big(\alpha(\proj{00} + \proj{11}) + (1-\alpha)(\proj{10} + \proj{01})\big)$
when computing the quantum capacity of $\CN_p$.  Furthermore, since $\CN_p$ is also \emph{degradable} \cite{ds}, $I_c(\rho,\CN_p)$ is concave as a function of $\rho$ (see Section~\ref{section:appendix:concavity} for a simple proof). 
Since $I_c(\rho_{\al},\CN_p) = I_c(\rho_{1-\al},\CN_p)$, one then concludes by symmetry that the maximum is achieved for $\al=\frac{1}{2}$, in which case $\rho_{\frac{1}{2}}$ is maximally mixed, yielding $\max_\rho I_c(\rho,\CN_p) = I_c(\pi^{A'B'},\CN_p) = 2-H(p)$. Note that this maximizing input is a product state, since $\pi^{A'B'} = \pi^{A'}\otimes\pi^{B'}$.
Define the Bell states
\begin{eqnarray*}
\ket{\psi_\pm}=\frac{1}{\sqrt{2}}\Big(\ket{00} \pm \ket{11}\Big).
\end{eqnarray*}
As $\ket{\psi_+}$ purifies the maximally mixed state $\pi_2$, let us write the global state 
\[\omega^{ABC} = \CN(\psi_+^{AA'}\otimes\psi_+^{BB'}).\]
Identifying $C=\h{A}\h{B}$ in the obvious way, we reexpress 
\[\omega^{A\h{A}B\h{B}} 
= (1-p)\psi_+^{A\h{A}}\otimes\psi_+^{B\h{B}} + p \psi_-^{A\h{A}}\otimes\psi_-^{B\h{B}}.\]
It is now a simple task to calculate 
\begin{eqnarray*}
H(ABC) &=& H(\omega) = H(p) \\
H(C) &=& H(\pi^{C}) = 2 \\
H(AC) &=& H(A\h{A}) + H(\h{B}) = H(p) + 1 = H(BC). 
\end{eqnarray*}
Combining these gives the relevant coherent informations
\begin{eqnarray*}
I_c(AB\,\rangle C) &=& H(C) - H(ABC) = 2 - H(p) \\
I_c(A\,\rangle BC) &=& H(BC) - H(ABC) = 1 + H(p) -H(p) = 1 \\
I_c(B\,\rangle AC) &=& H(AC) - H(ABC) = 1.
\end{eqnarray*}
Clearly $I_c(A\,\rangle BC)\leq \log|A'| = 1$ and $I_c(B\,\rangle AC)\leq \log|B'| = 1$ for any state arising from $\CN_p$. The individual rate bounds are thus saturated and the claim follows.
\end{proof}

\subsection{Concavity of coherent information for degradable channels}
\label{section:appendix:concavity}
Recall that a given a channel $\CN\colon A'\rightarrow B$, one may define a complementary channel $\CN_c\colon A'\rightarrow E$ as $\tr_B\CU$,  where $\CU\colon A'\rightarrow BE$ is any isometric extension of $\CN$.  The channel $\CN$ is then said to be \emph{degradable} \cite{ds} if there exists a channel $\CD\colon B\rightarrow E$ which degrades $\CN$ to $\CN_c$, in the sense that $\CN_c = \CD\circ\CN$.  A single-letter description of the quantum capacity of any such channel was given in $\cite{ds}$ as $\max_\rho I_c(\rho,\CN)$. 
We will show that if $\CN$ is degradable, then $I_c(\rho,\CN)$ is a concave function of $\rho$.
\begin{proof}
Fixing density matrices $\rho_0^{A'}$ and $\rho_1^{A'}$, we write $\rho^{UA'} = \lambda_0 \proj{0}^U \otimes\rho_0 + \lambda_1 \proj{1}^U\otimes\rho_1$, where $0\leq \lambda_0,\lambda_1$ satisfy $\lambda_0 + \lambda_1 = 1$.
Setting $\sig^{UBE} = \CU(\rho^{UA'})$ and $\sig'^{UE'F} = \CV(\tr_E\sig)$, where $\CV\colon B\rightarrow E'F$ isometrically extends the associated degrading map $\CD$, we write 
\begin{eqnarray*}
\lambda_0I_c(\rho_0,\CN) + \lambda_1I_c(\rho_1,\CN) &=& H(B|U)_\sig - H(E|U)_\sig \\
&=& H(FE'|U)_{\sig'} - H(E'|U)_{\sig'} \\
&=& H(F|E'U)_{\sig'} \\
&\leq& H(F|E')_{\sig'} \\
&=& H(B)_\sig - H(E)_\sig \\
&=& I_c(\lambda_0\rho_0 + \lambda_1\rho_1,\CN).
\end{eqnarray*}
Here, we used the fact that $\CV$ preserves entropies and the form of strong subadditivity which states that conditioning cannot increase entropy.  This proves the claim.
\end{proof}

\subsection{Proof of convexity of $\CC\CQ$ and $\CQ$}
\label{section:appendix:convex}
Let $\CN:A'B'\rightarrow C$ be a quantum multiple access channel.  We will prove that
$\CQ(\CN)$ is convex, as the proof for $\CC\CQ$ is identical.
\begin{proof}
Let $k_0$ and $k_1$ be positive integers, and fix any two states of the form (\ref{th2arise}), $\sigma_0^{A_0B_0C^{k_0}}$ and $\sigma_1^{A_1B_1C^{k_1}}.$
Then $(R_0,S_0),(R_1,S_1)\in \CQ(\CN)$, where for $i\in\{0,1\}$,
\begin{eqnarray*}
R_i &=& \frac{1}{k_i}I_c(A_i\,\rangle C^{k_i})_{\sigma_i} \\
S_i &=& \frac{1}{k_i}I_c(B_i\,\rangle C^{k_i})_{\sigma_i}.
\end{eqnarray*}
We will now show that for any rational $0\leq \lambda \leq 1$, $\lambda(R_0,S_0) + (1-\lambda)(R_1,S_1) \in \CQ(\CN).$  We first write $\lambda = \frac{\alpha}{\beta},$
for integers satisfying $\beta> 0,$ $\beta\geq \alpha \geq 0$. Setting $p_0=\alpha k_1,$ $p_1 = (\beta-\alpha)k_0,$ and  
$k = p_0k_0 + p_1k_1$,
define the composite systems $A = A_0^{p_0}A_1^{p_1}$ and 
$B = B_0^{p_0}B_1^{p_1}$,
as well as the density matrix
$\sigma^{ABC^k} = \sigma_0^{\otimes p_0}\otimes\sigma_1^{\otimes p_1},$ which is also of the form (\ref{th2arise}).
Additivity of coherent information across product states and some simple algebra gives
\begin{eqnarray*}
\frac{1}{k}I_c(A\,\rangle C^k)_{\sigma} &=&
\frac{p_0}{k}I_c(A_0\,\rangle C^{k_0})_{\sigma_0}+
\frac{p_1}{k}I_c(A_1\,\rangle C^{k_1})_{\sigma_1} \\
&=& \frac{p_0k_0 R_0 + p_1k_1R_1}{p_0k_0 + p_1k_1} \\
&=& \lambda R_0 + (1-\lambda) R_1.
\end{eqnarray*}
An identical calculation shows that
$\frac{1}{k}I_c(B\,\rangle C^k)_{\sigma} = \lambda S_0 + (1-\lambda) S_1.$
As $\CQ(\CN)$ was defined as the topological closure of rate pairs corresponding to states which appropriately arise from the channel, the result follows because the set of previously considered $\lambda$'s comprises a dense subset of the unit interval.
\end{proof}

\subsection{Proof of cardinality bound on $\CX$.}
\label{section:appendix:cardinality}
Begin by fixing a finite set $\CX$, a labelled collection of pure states $\{\ket{\phi_x}^{A'}\}_{x\in\CX}$, and a pure bipartite state $\ket{\Psi}^{BB'}.$ For each $x$, these define the states
$\sigma^{BC}_x = \CN(\phi_x\otimes\Psi)$ and $\omega^C_x = \tr_B\sigma_x$.  Assume for now that $|A'| \geq |C|$. Define a mapping $f\colon\CX\rightarrow \mathbb{R}^{|C|^2 + 1}$,
via
\[f \colon x \mapsto f_x\equiv(\omega_x,H(\omega_x),I_c(B\,\rangle C)_{\sigma_x}),\]
where we identify $\omega_x$ with its $|C|^2-1$ dimensional parameterization.
By linearity, this extends to a map from probability mass functions on $\CX$ to $\mathbb{R}^{|C|^2 + 1},$ where
\[f\colon p(x)\mapsto \sum_x p(x)f_x \equiv (\omega_p, H(C|X)_p,I_c(B\,\rangle CX)_p),\]
Our use of the subscript $p$ should be clear from the context.
The use of Caratheodory's theorem for bounding the support sizes of auxilliary
random variables in information theory (see \cite{ck}) is well-known.  Perhaps less familiar is the observation \cite{wynerziv,salehiaux} that a better bound can often be obtained by use of a related theorem by Fenchel and Eggleston \cite{egg}, which states that if $S\subseteq\mathbb{R}^n$ is the union of at most $n$ connected subsets, and if $y$ is contained in the convex hull of $S$, then $y$ is also contained in the convex hull of at most $n$ points in $S$.  As the map $f$ is linear, it maps the simplex of distributions on $\CX$ into a single connected subset of $\mathbb{R}^{|C|^2+1}$.  Thus, for any distribution $p(x)$, there is another distribution $p'(x)$ which puts positive probability on at most $|C|^2+1$ states, while satisfying $f(p) = f(p').$  If it is instead the case that $|A'|< |C|,$ this bound can be reduced to $|A'|^2 + 1$ by replacing the first components of the map $f$ with a parameterization of $\phi_x^{A'}$, as specification of a density matrix on
$A'$ is enough to completely describe the resulting state on $C$.  It is therefore sufficient to consider $|X| \leq \min\{|A'|,|C|\}^2 + 1$ in computing $\CC\CQ^{(1)}(\CN)$.

\qed

\end{document}